\documentclass[preprint,5p,authoryear]{elsarticle}

\usepackage{lineno,hyperref}

\usepackage{multicol}
\usepackage{booktabs}
\usepackage{bm}
\usepackage{xcolor}
\usepackage{threeparttable}
\usepackage{caption}
\usepackage[ruled,vlined,linesnumbered]{algorithm2e}
\usepackage{paracol}
\globalcounter{algocf}

\SetCommentSty{mycommfont}
\SetNlSty{bfseries}{\color{black}}{}
\SetArgSty{textnormal}
\DontPrintSemicolon
\let\oldnl\nl
\newcommand{\nonl}{\renewcommand{\nl}{\let\nl\oldnl}}

\usepackage{physics}
\DeclareDocumentCommand\vectorbold{ s m }{\IfBooleanTF{#1}{\boldsymbol{#2}}{\mathbf{#2}}} 
\DeclareDocumentCommand\vb{}{\vectorbold} 

\modulolinenumbers[5]

\journal{Journal of \LaTeX\ Templates}

\begin{document}

\begin{frontmatter}

\title{The Gaia AVU--GSR parallel solver: preliminary studies of a LSQR--based application in perspective of exascale systems}

 \author[1]{Valentina Cesare\corref{cor1}}
	\ead{valentina.cesare@inaf.it}
	\author[1]{Ugo Becciani}
	\ead{ugo.becciani@inaf.it}
	\author[2]{Alberto Vecchiato}
	\ead{alberto.vecchiato@inaf.it}
	\author[2]{Mario Gilberto Lattanzi}
	\ead{mario.lattanzi@inaf.it}
	\author[3]{Fabio Pitari}
	\ead{f.pitari@cineca.it}
	\author[1]{Mario Raciti}
	\ead{mario.raciti@inaf.it}
	\author[1]{Giuseppe Tudisco}
	\ead{giuseppe.tudisco@inaf.it}
	\author[4]{Marco Aldinucci}
	\ead{marco.aldinucci@unito.it}
	\author[2]{Beatrice Bucciarelli}
	\ead{beatrice.bucciarelli@inaf.it}
	
\cortext[cor1]{Corresponding author}
\address[1]{INAF, Astrophysical Observatory of Catania, via Santa Sofia 78, 95123 Catania, CT, Italy}
\address[2]{INAF, Astrophysical Observatory of Turin, via Osservatorio 20, 10025 Pino Torinese, TO, Italy}
\address[3]{CINECA, via Magnanelli 6/3, 40033 Casalecchio di Reno, BO, Italy}
\address[4]{University of Turin, Computer Science Department, corso Svizzera 185, 10149 Turin, TO, Italy}

 
 
 
 

\begin{abstract}
The Gaia Astrometric Verification Unit--Global Sphere Reconstruction (AVU--GSR) Parallel Solver aims to find the astrometric parameters for $\sim$10$^8$ stars in the Milky Way, the attitude and the instrumental specifications of the Gaia satellite, and the global parameter $\gamma$ of the post Newtonian formalism. The code iteratively solves a system of linear equations, $\mathbf{A} \times \vb*{x} = \vb*{b}$, where the coefficient matrix $\mathbf{A}$ is large ($\sim$$10^{11} \times 10^8$ elements) and sparse. To solve this system of equations, the code exploits a hybrid implementation of the iterative PC-LSQR algorithm, where the computation related to different horizontal portions of the coefficient matrix is assigned to {separate} MPI processes. In the original code, each matrix portion is further parallelized over the OpenMP threads. To further improve the code performance, we ported the application to the GPU, replacing the OpenMP parallelization language with OpenACC. In this port, $\sim$95\% of the data is copied from the host to the device at the beginning of the entire cycle of iterations, making the code \textit{compute bound} rather than \textit{data-transfer bound}. The OpenACC code presents a speedup of $\sim$1.5 over the OpenMP version but further optimizations are in progress to obtain higher gains. The code runs on multiple GPUs and it was tested on the CINECA supercomputer Marconi100, in anticipation of a {port} to the pre-exascale system Leonardo, that will be installed at CINECA in 2022. 
\end{abstract}


\end{frontmatter}


\section{Introduction}
\label{sec:Intro}

The ESA's Gaia mission\footnote{\url{https://sci.esa.int/web/gaia}} has provided, in the eight years since its launch on 19$^{\rm th}$ December 2013, a map both in the position and in the velocity dimensions of $\sim$1 billion of stars in the Milky Way, about 1\% of its total content, with micro-arcsecond accuracy. 
Two of the main objectives of the mission are the investigation of the formation and the evolution of our galaxy (e.g.~\citealt{Krolikowski_2021}) and the test of Einstein's theory of General Relativity (GR) (e.g.~\citealt{Hees_2018}). Indeed, thanks to the accuracy of the measurements, Gaia can detect the bending of the light around massive objects and the consistency of this effect with the predictions of GR can be verified. 

In the Gaia's early third data release (EDR3)~\citep{Gaia_EDR3_2021} a catalogue of parallaxes, sky positions, and proper motions of $\sim$1.468$\times 10^9$ stars was published and the complete Gaia's third data release (DR3) {was published in June 2022}. 
The application that we present in this work, called Gaia Astrometric Verification Unit--Global Sphere Reconstruction (AVU--GSR) Parallel Solver, is developed under the Data Processing and Analysis Consortium (DPAC), and it aims to find these parameters for the so-called \textit{primary stars} of the global astrometric sphere of the Gaia mission, comprising $\sim$10$^8$ stars. Moreover, it will constrain the attitude and the instrumental settings of the Gaia satellite and the parameter $\gamma$ of the Parametrized Post-Newtonian (PPN) formalism of relativistic gravity theories
to describe the space-time and to test GR against alternative theories of gravity~\citep{Vecchiato_2003}. 

The astrometric model of the observations produces a set of non-linear equations, one for each observation of the Gaia satellite, as a function of the unknowns. To render this system of equations computationally solvable, the observation equations are linearized around an appropriate starting point. The resulting linearized system can be written in matrix form as: 
\begin{equation}
\label{eq:Axb}
\mathbf{A} \times \vb*{x} = \vb*{b},
\end{equation}
where $\mathbf{A}$ is the coefficient matrix, $\vb*{x}$ the vector of the unknowns, and $\vb*{b}$ the vector of the known terms. The system matrix $\mathbf{A}$ is large and sparse ($\sim$$10^{11} \times 10^8$ elements), where the number of rows is the number of observations of the stellar parameters and the number of columns is the number of unknowns to solve. During computation only the non-zero elements of $\mathbf{A}$ are considered, and the new matrix contains $\sim$$10^{11} \times 10^1$ elements. 
A single star is observed $\sim$10$^3$ times, on average. Since the number of equations is much larger than the number of unknowns, the system has to be solved in the least-squares sense, adopting a modified version of the iterative LSQR algorithm~\citep{Paige_and_Saunders_1982a,Paige_and_Saunders_1982b}, where the iterations stop when either a given convergence criterion or a maximum number of iterations is achieved. 

The latest version of the code is written in C/C++ and is hybridly parallelized with MPI + OpenMP~\citep{Becciani_2014}. Each MPI process deals with the computation of a horizontal portion of the coefficient matrix, a subset of the total number of observations, and the calculation in each MPI process is further parallelized over the OpenMP threads. 

{Here we present a preliminary port of this application to GPUs, where we replace the OpenMP part with the parallelization language OpenACC, finalized to a more optimized port with CUDA to pre-exascale systems.}
The OpenACC code runs on multiple GPUs and currently presents a moderate gain in performance over the OpenMP version, which might be improved with the future optimizations both with OpenACC and CUDA. We performed the performance tests on the MPI + OpenMP and MPI + OpenACC versions of the application on the CINECA supercomputer Marconi100 (M100). 

The contents of the paper are as follows: Section~\ref{sec:Related_works} briefly describes the usage of the LSQR algorithm in the literature; Section~\ref{sec:Coeff_matrix} presents the structure of the coefficient matrix of the system of equations; Section~\ref{sec:Parallel_MPI_OMP} details the present version of the code parallelized with MPI + OpenMP, in production on M100; Section~\ref{sec:Production} describes the platform on which the code that is in production runs and presents the performance of a typical execution of the in-production code; Section~\ref{sec:GPU} presents {the current port to the GPU} with MPI + OpenACC; Section~\ref{sec:Performance_tests} describes the tests, performed on M100, that compare the performance of the MPI + OpenMP and MPI + OpenACC applications; Section~\ref{sec:Conclusions_and_future_works} concludes the paper, also introducing the future work with CUDA aimed to further accelerate the code.

\section{Related work}
\label{sec:Related_works}


The main algorithm of this application is based on the LSQR, an iterative Krylov subspace algorithm conceived to solve large scale ill-posed problems while maintaining numerical stability~\citep{Paige_and_Saunders_1982a,Paige_and_Saunders_1982b}. Typically, the LSQR algorithm is employed to solve, in the least-squares sense, a system of equations with a large and sparse coefficient matrix that has not a unique solution (e.g.~\citealt{Paige_and_Saunders_1982a,Paige_and_Saunders_1982b,Reichel_and_Ye_2008,Jaffri_2020,Penghui_2020}). 
To mention some examples, this algorithm is exploited in the following contexts: (I) geophysics, to locate underground gravitational and magnetic anomalies~\citep{Joulidehsar_2018,LIANG_2019,LSQR_geology_2019}; (II) medicine, in electrocardiography ~\citep{Bin_2020} and $X$-ray tomography~\citep{Guo_2021}; (III) industry, in electrical resistance tomography~\citep{Jaffri_2020}; (IV) astronomy, for radioastronomical image reconstruction~\citep{Naghibzadeh_and_vanderVeen_2017}, and in the application presented in this paper. 

The LSQR algorithm was also adopted for the data reduction of the High Precision Parallax Collecting Satellite (Hipparcos) Space Astrometry Mission, the first astrometric space mission defined within the ESA scientific program and precursor of the Gaia mission~\citep{Borriello_1986,VanderMarel_1988,Becciani_2014}.
In several research fields, this algorithm is used to solve an \textit{inverse problem}, one of the most important issues in mathematical sciences, that consists of estimating the parameters of a model from a set of observational data (e.g.~\citealt{LSQR_geology_2019,Bin_2020,Guo_2021}).  
This is also the target of the Gaia AVU--GSR application.

The LSQR algorithm usually adopts a \textit{preconditioning technique} (e.g.~\citealt{Ling_2019,Bin_2020}) to improve its convergence speed, that could consist of properly normalizing the coefficient matrix of the system before starting the iterations and multiplying this normalization factor with the solution obtained only at the end of the computation. This allows the LSQR algorithm to find an equally accurate solution in $\sim$60-70\% of the time compared to other standard algorithms such as the conventional iterative reweight norm method~\citep{Bin_2020}.  In the Gaia AVU-GSR code, we preconditioned the system of equations by dividing the parameters of each column of the coefficients matrix by the norm of the column itself. We stored these normalization factors in an array $\vb*{p}$ with a number of elements equal to the number of columns of the coefficient matrix and we {multiplied} this array {with} the solution and standard error arrays at the end of all the iterations. The preconditioned LSQR algorithm can be abbreviated as PC-LSQR.

Typically, the systems of equations that exploit the LSQR algorithm require parallelism to be solved in reasonable timescales and to overcome possible problems due to memory limits, given the large size of their coefficient matrices~\citep{Huang_2012}. There are several existing implementations of the parallel LSQR algorithm. For example, the implementation of~\citet{Baur_and_Austen_2005} exploits repeated vector-vector operations and was applied to the data taken by the CHAMP, GRACE, and GOCE satellites, and the implementations of~\citet{Liu_2006} and~\citet{Huang_2013} {were} employed in seismic tomography. Another example {is provided by} the Portable, Extensible Toolkit for Scientific Computation (PETSc) libraries~\citep{petsc-efficient,petsc-user-ref,petsc-web-page}, optimized scientific libraries that support MPI and GPU (CUDA and OpenCL) parallelism and even hybrid MPI + GPU parallelization paradigms. 

The main issue for these libraries is that they are not optimized for sparse matrices with particular patterns of non-zero elements~\citep{Yoo_2011} and, thus, they are not suitable to solve {systems} of equations, {such as} the one of the Gaia AVU--GSR application, that {are} based on a very peculiar sparsity scheme, as described in Section~\ref{sec:Coeff_matrix}.
When dealing with such large systems, optimization is essential: assuming a system that converges in 141000 iterations, with $4.23$~s per iteration, {representative of a typical execution of the Gaia AVU-GSR code (see Section~\ref{sec:Production})}, saving 1~s per iteration means saving 2350 hours, {that is} the $\sim$24\% of the initial execution time. These percentages might vary for systems with different sizes. For this reason, we did not adopt an existing implementation of the LSQR algorithm but defined a customized implementation, which exploits a preconditioning technique and an ad-hoc algorithm to compress the sparse coefficient matrix of the system of equations.

We ported our application to a GPU environment, using the OpenACC parallel programming model. Some implementations of GPU-ported LSQR algorithm were created for solving problems in specific fields, e.g. seismic~\citep{Huang_2012} or medical~\citep{Flores_2016} tomography. In particular,~\citet{Huang_2012} implemented a MPI--CUDA version of the LSQR algorithm with excellent performance: 17.6x speedup over the serial CPU algorithm and 3.8x speedup over the MPI--CPU algorithm, besides obtaining a better performance than the PETSc implementation and very good strong and weak scaling behaviours\footnote{The \textit{strong scaling} is the capability of a parallel software to compute a fixed-size problem faster with more computing resources~\citep{Amdahl_1967}. Instead, the \textit{weak scaling} is the capability of a parallel software to maintain constant the computation time of a problem with a size proportional to the amount of computing resources~\citep{Gustafson_1988}.}. 

\section{Coefficient matrix structure}
\label{sec:Coeff_matrix}

As mentioned in Section~\ref{sec:Intro}, this application solves a system of linear equations given by Eq.~\eqref{eq:Axb}, where the coefficient matrix $\mathbf{A}$ contains $\sim$10$^{11} \times 10^8$ elements. To solve the system, we may choose between two fully-relativistic astrometric models \citep{Bertone_2017,Crosta_2017}, whose unknowns are~\citep{Becciani_2014}:
\begin{enumerate}
	\item the astrometric unknowns, {i.e.} the parallax $\omega$, the right ascension $\alpha$, the declination $\delta$, and the two components of the proper motion along these two directions, $(\mu_\alpha,\mu_\delta)$, of every star;
	\item the attitude unknowns, given by a proper B-spline representation of the Rodrigues parameters that describe the satellite attitude during the entire duration of the mission;
	\item the instrumental unknowns;
	\item the global unknown $\gamma$ of the PPN formalism;
\end{enumerate}
with a total of $\sim$10$^8$ unknowns.

To solve the system of equations, we adopted a hybrid implementation of PC-LSQR, an iterative conjugate-gradient type algorithm that can solve an overdetermined system of linear equations in the least-squares sense. The most computationally demanding part of this algorithm, and of the entire application, consists of the call, at each iteration $i$, of the \textit{aprod} function in one of two possible modes. In the mode 1, aprod computes the product $\mathbf{A} \times \vb*{x}^i$, where $\vb*{x}^i$ is the $i$-th approximation of the solution of the system. In the mode 2, aprod computes the product $\mathbf{A}^T \times \vb*{y}^i$, where $\vb*{y}^i$ is proportional to the iterative estimation of the vector of the residuals, $\vb*{y'}^i = \vb*{b} - \mathbf{A} \times \vb*{x}^i$, and $\vb*{b}$ is the vector of the known terms. The vector of the residuals is the quantity that has to be minimized in the least-squares sense up to the convergence of the algorithm. In fact, the convergence is considered achieved when $\vb*{y'}^i$ goes below a predefined tolerance $tol$, {set to the machine precision ($10^{-16}$ on M100). To find a unique solution, an additional number of constraint equations have to be set.}

\begin{figure}[hbt!]
	\centering
	\includegraphics[width=0.33\textwidth]{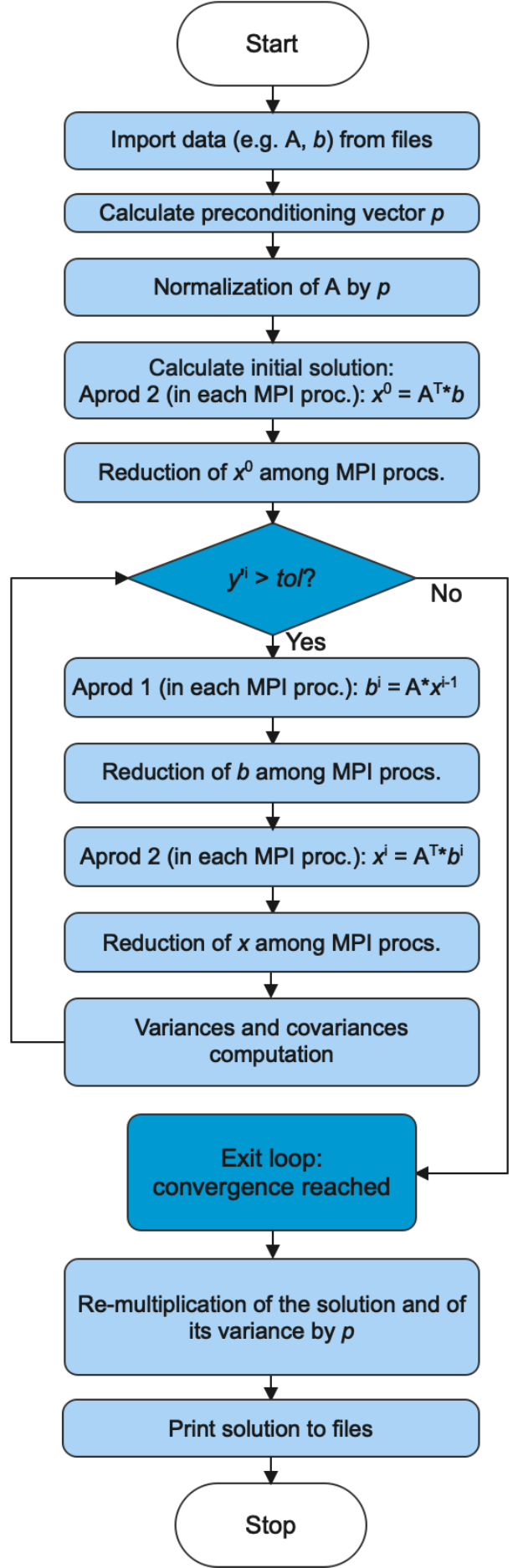}
	\caption{Flowchart of the entire Gaia AVU--GSR application.}
	\label{fig:Flowchart_all}
\end{figure}

{Figure~\ref{fig:Flowchart_all} shows the flowchart of the entire code. First of all, the quantities employed in the calculations, such as the coefficient matrix $\mathbf{A}$ and the known terms array $\vb*{b}$, are imported from external binary files, converted in binary format from original FITS files with an external program. Then, we calculate the preconditioning array $\vb*{p}$ 
and we normalize each column of $\mathbf{A}$ by the elements of this vector (Section~\ref{sec:Related_works}). 
From the normalized coefficient matrix and the known terms array, we calculate, in each MPI process, the initial solution of the system of equations, with the aprod 2 function. The initial solution is then reduced among the different MPI processes. Then, the LSQR algorithm starts. The LSQR algorithm consists of a while loop that terminates when either the convergence condition or, in case convergence cannot be achieved, a maximum number of iterations 
are reached. 
At each iteration $i$ of the LSQR algorithm, the aprod 1 and 2 functions are called to provide the iterative estimates of the known terms $\vb*{b}$ and of the unknowns $\vb*{x}$, respectively, locally to each MPI process. The $\vb*{b}$ and the $\vb*{x}$ arrays are then reduced among the MPI processes. At the end of each iteration, we also compute the standard errors on the unknowns (variances) and the possible correlations between the unknowns (covariances). After the end of the LSQR cycle, we re-multiply the solution and the error on the solution by the preconditioning vector $\vb*{p}$. The algorithm concludes with the print of the solution, its standard error, and the covariances to binary files, that are converted to FITS format with an external program.}

\begin{figure*}[hbt!]
	\centering
	\includegraphics[width=0.95\textwidth]{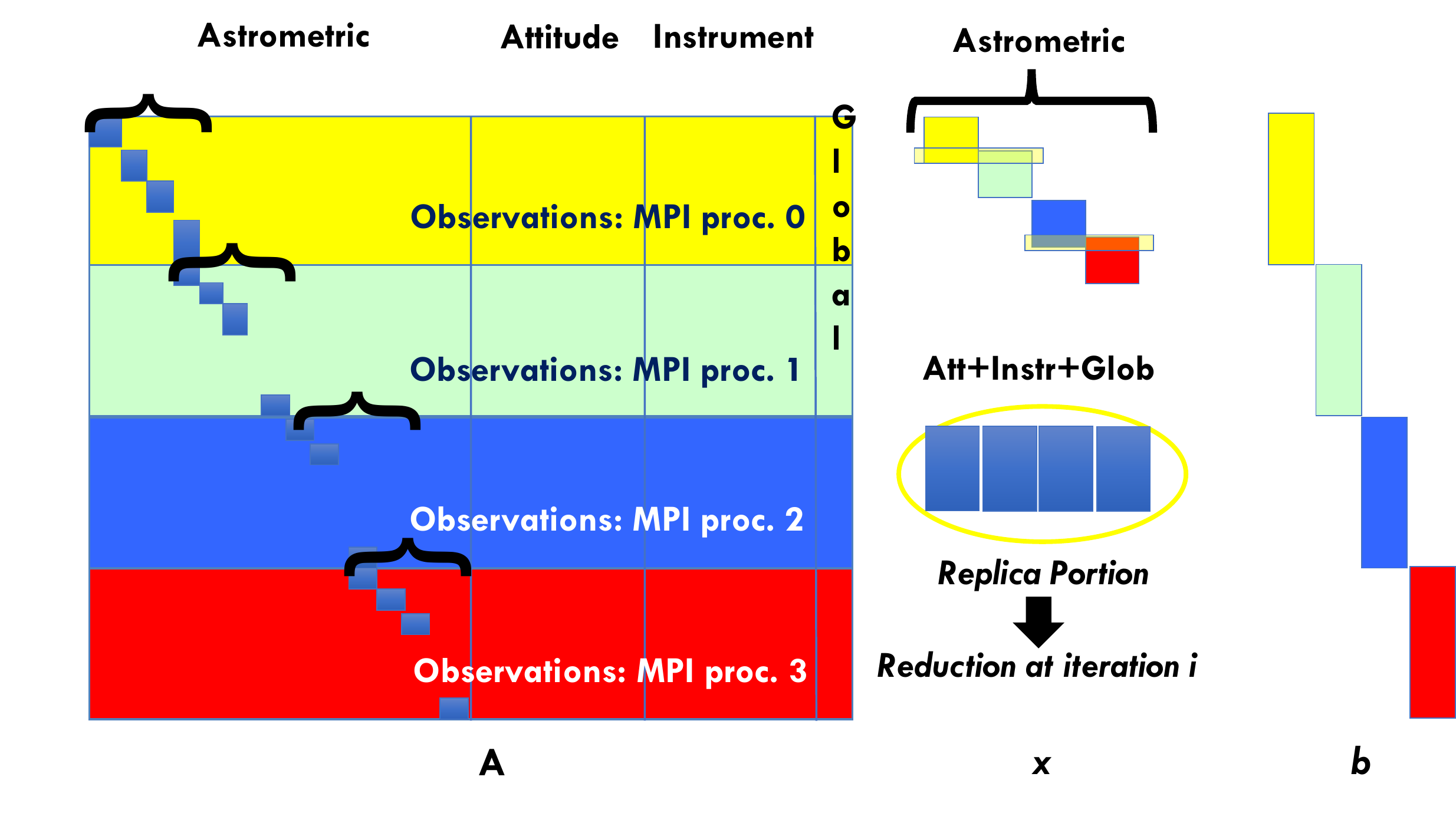}
	\caption{Parallelization scheme of the system of equations (Eq.~\ref{eq:Axb}) on four MPI processes in a single node of a computer cluster. {From the left to the right, we can see the coefficient matrix $\mathbf{A}$, the unknowns vector $\vb*{x}$, and the known terms vector $\vb*{b}$.} The coloured stripes refer to the four MPI processes where the computation is performed. The block-diagonal part in the left side of the coefficient matrix represents its astrometric section. {The four square blocks diagonally arranged, colour-coded as the MPI processes, and labelled as ``Astrometric'' show the astrometric part of the solution array, distributed among the MPI processes.} Instead, the four blue aligned blocks, labelled as ``Att+Instr+Glob'', represent the attitude, instrumental, and global portions of the solution array, replicated on each MPI process, as written below. At the end of each iteration $i$, a reduction of the replicated portions of $\vb*{x}$ is performed.}
	\label{fig:Par_Data_schemes}
\end{figure*}

Each observation equation (each row of the coefficient matrix $\mathbf{A}$) contains the astrometric, the attitude, the instrumental, and the global parameters {(vertical stripes on the coefficient matrix in Figure~\ref{fig:Par_Data_schemes})}.  
{The astrometric coefficients represent the $\sim$90\% of $\mathbf{A}$ and they are equal to $N_{\rm Astro} \times N_{\rm Stars} \times N_{\rm Obs}$, where $N_{\rm Astro} = 5$ is the number of astrometric parameters per star, $N_{\rm Stars} \in [10^6,10^8]$ is the number of stars, and $N_{\rm Obs} \sim 10^{11}$ is the number of observations.}
{The non-zero astrometric coefficients of $\mathbf{A}$ are} organized in a block-diagonal structure {(left part of the coefficient matrix in Figure~\ref{fig:Par_Data_schemes})}, where the rows of each block contain the astrometric parameters observed for a certain star. Specifically, the first block is related to star number 0, the second block to star number 1, and so forth.

After the astrometric section, each row of the attitude part of the matrix $\mathbf{A}$ contains $N_{\rm Att}=12$ parameters different from zero, organized in $N_{\rm Axes} = 3$ blocks of $N_{\rm ParAxis}=4$ elements separated by $N_{\rm DFA}$ elements equal to zero. $N_{\rm Axes}$ represents the number of axes of the satellite attitude, in this case equal to 3, whereas $N_{\rm DFA}$ is the number of degrees of freedom carried by each of the three axes. After the attitude section, each row of the instrumental part of the matrix $\mathbf{A}$ contains $N_{\rm Instr} = 6$ elements different from zero, distributed without any predefined order, and the last columns of $\mathbf{A}$ contain the coefficients of the global unknowns. In our case, we only considered $N_{\rm Glob} = 1$ global parameter, {which is} the parameter $\gamma$ of the PPN formalism. {The left part of Figure~\ref{fig:Par_Data_schemes} summarizes the structure of the system matrix $\mathbf{A}$.}

The coefficient matrix $\mathbf{A}$ is large and sparse, {since} the ratio between the number of elements equal to zero and different from zero is much larger than one. With a large matrix of $\sim$$10^{11}\times 10^8$ elements, {solving the system could not be performed on any existing infrastructure, even with an efficient parallelization scheme.}  
{For this reason, the matrix is compressed through an ad-hoc algorithm~\citep{Becciani_2014} } and only the non-zero elements of $\mathbf{A}$ are considered during computation.
In this way, the matrix size reduces from $\sim$$10^{11}\times 10^8$ to $\sim$$10^{11}\times 10^1$. The {new} matrix, $\mathbf{A}_\text{\bf d}$, {where ``d'' stands for ``dense'',} is stored in a 1D double-precision array that contains, for each observation, $N_{\rm par} = 24$ non-zero parameters ($N_{\rm Astro} = 5$ astrometric coefficients, $N_{\rm Att} = 12$ attitude coefficients, $N_{\rm Instr} = 6$ instrumental coefficients, and $N_{\rm Glob} = 1$ global coefficient).

{To solve the system of equations with $\mathbf{A}_\text{\bf d}$,} we defined a map between the indexes of the elements in $\mathbf{A}_\text{\bf d}$ and in the original matrix $\mathbf{A}$. The 1D single-precision array $\vb*{M}_\text{\bf i}$ contains the number of the star and the index of the first attitude element in the matrix $\mathbf{A}$, for each observation, i.e. for each row of $\mathbf{A}$. Given the regular structures of the astrometric and of the attitude sections in the matrix $\mathbf{A}$, described above, these two indexes are sufficient to retrieve the original positions of all the astrometric and the attitude coefficients in $\mathbf{A}$. Instead, the 1D single-precision array $\vb*{I}_\text{\bf c}$ contains the indexes pointing to the positions that all the instrumental elements of $\mathbf{A}_\text{\bf d}$ had in $\mathbf{A}$, since the instrumental part does not follow a regular pattern. Instead, the unknowns and the known terms are stored in two 1D double-precision arrays, $\vb*{x}$ and $\vb*{b}$, respectively.

\section{Parallel code structure: MPI + OpenMP}
\label{sec:Parallel_MPI_OMP}

The application is parallelized with MPI + OpenMP. Figure~\ref{fig:Par_Data_schemes} summarizes the parallelization scheme of the system of equations (Eq.~\ref{eq:Axb}) on four MPI processes, represented with four different colours, in a single node of a computer cluster. 

The computation associated with different horizontal portions of the matrix $\mathbf{A}_\text{\bf d}$, {i.e. with} different subsets of the total number of observations, is assigned to separate MPI processes. The number of observations handled by each MPI process is stored in a 1D single-precision array, $\vb*{N}[nproc]$, where $nproc$ is the total number of MPI processes on which the code runs. 

To optimize the code, the data are distributed among the MPI processes as evenly as possible. To perform this task, a small portion of the data, i.e. the attitude, the instrumental, and the global portions of the unknown vector $\vb*{x}$, is replicated on each MPI process. The ``Att+Instr+Glob -- Replica portion'' part {in Figure~\ref{fig:Par_Data_schemes}} represents this replicated portion of the solution array. The majority of the data, i.e. the astrometric section of $\vb*{x}$, is distributed, as we can see from the four ``Astrometric'' blocks {in Figure~\ref{fig:Par_Data_schemes}}, colour-coded as the four MPI processes. This scheme reduces the number of communications between the different MPI processes. At every iteration of the LSQR algorithm, a reduce operation is performed to combine the results, {i.e.} the replicated parts of the $i$-th iterative estimates of both the known terms and the solution arrays, $\vb*{b}^i$ and $\vb*{x}^i$, from the different MPI processes (Figures~\ref{fig:Flowchart_all} and~\ref{fig:Par_Data_schemes}).

The reason for distributing the astrometric part and replicating the attitude, instrumental, and global parts on the MPI processes instead of distributing the entire system is due to the different arrangements of the coefficients in the four parts of the system. The coefficients of the astrometric section are regularly distributed in the system matrix following a block-diagonal structure, which makes it easy to assign their computation to different MPI processes. Instead, the other three sections do not follow a specific regular pattern, which makes more difficult to rearrange the code to distribute their computation among the MPI processes. Replicating the computation of the attitude + instrumental + global parts on the MPI processes simplifies the writing of the code while implying only a minor {loss} in performance, since these sections represent only the $\sim$10\% of the total.

The observations treated by each MPI process are further parallelized over the OpenMP threads, where the index of the thread, $tid$, goes from 0 to $nth - 1$, where $nth$ is the total number of threads set at runtime (see lines omp.\ref{algo:omp:aprod2_tid} and omp.\ref{algo:omp:aprod2_nth} of Algorithm~\ref{algo:aprod_2}). In the aprod 1, the parallelization with OpenMP is managed with the \texttt{\#pragma omp for} directive, placed on the for loop at line omp.\ref{algo:omp:N_pid} of Algorithm~\ref{algo:aprod_1}. In the aprod 2, we defined a single-precision 1D array, $\vb*{N}_\text{\bf t}$, that stores the number of observations dealt by each OpenMP thread $tid$, similarly to the array $\vb*{N}$ for each MPI process of rank $pid$. The for loop that manages the computation in each OpenMP thread of ID $tid$ iterates from $\vb*{N}_\text{\bf t}[tid][0]$ to $\vb*{N}_\text{\bf t}[tid][1]$, where $\vb*{N}_\text{\bf t}[tid][0]$ and $\vb*{N}_\text{\bf t}[tid][1]$ are the first and the last observation in the $tid$-th thread (see line omp.\ref{algo:omp:aprod2_Ntid} of Algorithm~\ref{algo:aprod_2}).

\begin{algorithm*}
	\begin{multicols}{2}
		\nonl \textbf{\small aprod 1 with OpenMP}\;
		\SetNlSty{}{omp.}{}
		\textcolor{blue}{\texttt{\#pragma omp parallel private($pid$,$sum$) shared($\vb*{N}$,$\vb*{x}$,$\mathbf{A}_\textbf{d}$,$\vb*{b}$)}} \;
		\{\;
		\text{ }\text{ }\text{ }\textcolor{blue}{\texttt{\#pragma omp for}}\;
		\text{ }\text{ }\text{ }\For{$i\leftarrow 0$ \KwTo $\vb*{N}[pid]$}  { \label{algo:omp:N_pid}
			$sum = 0.0$\;
			\tcp{Astrometric sect.}
			$k = i \times N_{\rm par}$ \;
			\For{$j\leftarrow 0$ \KwTo $N_{\rm Astro}$}{
				$sum = sum + \mathbf{A}_\textbf{d}[k]\vb*{x}[j + {\rm offset}[i]]$\;
				$k$++\;
			}
			\nonl \;
			\nonl \;
			\nonl \;
			\nonl \;
			\nonl \;
			\nonl \;
			\nonl \;
			\nonl \;
			\tcp{Attitude sect.}
			$k = i \times N_{\rm par} + N_{\rm Astro}$ \;
			\For{$j_1\leftarrow 0$ \KwTo $N_{\rm Axes}$}{
				$k_2 = j_1 \times N_{\rm DFA} + {\rm offset}[i] $\;
				\For{$j_2 \leftarrow 0$ \KwTo $N_{\rm ParAxis}$}{
					$sum = sum + \mathbf{A}_\textbf{d}[k]\vb*{x}[j_2 + k_2]$\;
					$k$++\;
				}  
			}
			\nonl \;
			\nonl \;
			\nonl \;
			\nonl \;
			\nonl \;  
			\nonl \;  
			\nonl \;
			\tcp{Instrumental sect.} 	        
			$k = i \times N_{\rm par} + N_{\rm Astro} + N_{\rm Att}$ \;
			\For{$j\leftarrow 0$ \KwTo $N_{\rm Instr}$}{
				$sum = sum + \mathbf{A}_\textbf{d}[k]\vb*{x}[\mathcal{F}(i,j) + {\rm offset}]$\;
				$k$++\;
			}
			\nonl \;
			\nonl \;
			\nonl \;
			\nonl \;
			\nonl \;
			\nonl \;  
			\nonl \;  
			\tcp{Global sect.}
			$k = i \times N_{\rm par} + N_{\rm Astro} + N_{\rm Att} + N_{\rm Instr}$ \;
			\For{$j\leftarrow 0$ \KwTo $N_{\rm Glob}$}{
				$sum = sum + \mathbf{A}_\textbf{d}[k]\vb*{x}[j + {\rm offset}]$\;
				$k$++\;
			}
			$\vb*{b}[i]$ = $\vb*{b}[i]$ + $sum$\; 
		}
		\}\;
		Constraints computation\;
		
		

		
		\setcounter{AlgoLine}{0}
		\SetNlSty{}{acc.}{}
		\nonl \textbf{\small aprod 1 with OpenACC}\;
		\tcp{Astrometric sect.}
		\textcolor{red}{\texttt{\#pragma acc parallel private($sum$)}} \; \label{algo:acc:astro_par_sect_beg}
		\{\;
		\text{ }\text{ }\text{ }\textcolor{red}{\texttt{\#pragma acc loop}} \; \label{algo:acc:astro_par_loop}
		\text{ }\text{ }\text{ }\For{$i\leftarrow 0$ \KwTo $\vb*{N}[pid]$}  {
			$sum = 0.0$\;
			$k = i \times N_{\rm par}$\;
			\For{$j\leftarrow 0$ \KwTo $N_{\rm Astro}$}{
				$sum = sum + \mathbf{A}_\textbf{d}[k + j]\vb*{x}[j + {\rm offset}[i]]$\;
				\nonl \; 
			}
			$\vb*{b}[i]$ = $\vb*{b}[i]$ + $sum$\; 
		}
		\}\; \label{algo:acc:astro_par_sect_end}
		\tcp{Attitude sect.}
		\textcolor{red}{\texttt{\#pragma acc parallel private($sum$)}} \; \label{algo:acc:att_par_sect_beg}
		\{\;
		\text{ }\text{ }\text{ }\textcolor{red}{\texttt{\#pragma acc loop}} \; \label{algo:acc:att_par_loop}
		\text{ }\text{ }\text{ }\For{$i\leftarrow 0$ \KwTo $\vb*{N}[pid]$}  {
			$sum = 0.0$\;
			$k = i \times N_{\rm par} + N_{\rm Astro}$\;
			\For{$j_1\leftarrow 0$ \KwTo $N_{\rm Axes}$}{
				$k_1 = j_1 \times N_{\rm ParAxis} $\;
				$k_2 = j_1 \times N_{\rm DFA} + {\rm offset}[i] $\;
				\For{$j_2 \leftarrow 0$ \KwTo $N_{\rm ParAxis}$}{
					$sum = sum + \mathbf{A}_\textbf{d}[k + j_2 + k_1]\vb*{x}[j_2 + k_2]$\;
				}  
			}
			$\vb*{b}[i]$ = $\vb*{b}[i]$ + $sum$\; 
		}
		\}\; \label{algo:acc:att_par_sect_end}
		\tcp{Instrumental sect.}
		\textcolor{red}{\texttt{\#pragma acc parallel private($sum$)}} \; \label{algo:acc:instr_par_sect_beg}
		\{\;
		\text{ }\text{ }\text{ }\textcolor{red}{\texttt{\#pragma acc loop}} \; \label{algo:acc:instr_par_loop}
		\text{ }\text{ }\text{ }\For{$i\leftarrow 0$ \KwTo $\vb*{N}[pid]$}  {
			$sum = 0.0$\;
			$k = i \times N_{\rm par} + N_{\rm Astro} + N_{\rm Att}$\;
			\For{$j\leftarrow 0$ \KwTo $N_{\rm Instr}$}{
				$sum = sum + \mathbf{A}_\textbf{d}[k + j]\vb*{x}[\mathcal{F}(i,j) + {\rm offset}]$\;
				\nonl \; 
			}
			$\vb*{b}[i]$ = $\vb*{b}[i]$ + $sum$\; 
		}
		\}\; \label{algo:acc:instr_par_sect_end}
		\tcp{Global sect.}
		\textcolor{red}{\texttt{\#pragma acc parallel private($sum$)}} \; \label{algo:acc:glob_par_sect_beg}
		\{\;
		\text{ }\text{ }\text{ }\textcolor{red}{\texttt{\#pragma acc loop}} \; \label{algo:acc:glob_par_loop}
		\text{ }\text{ }\text{ }\For{$i\leftarrow 0$ \KwTo $\vb*{N}[pid]$}  {
			$sum = 0.0$\;
			$k = i \times N_{\rm par} + N_{\rm Astro} + N_{\rm Att} + N_{\rm Instr}$\;
			\For{$j\leftarrow 0$ \KwTo $N_{\rm Glob}$}{
				$sum = sum + \mathbf{A}_\textbf{d}[k+j]\vb*{x}[j + {\rm offset}]$\;
				\nonl \; 
			}
			$\vb*{b}[i]$ = $\vb*{b}[i]$ + $sum$\;  
		}
		\}\; \label{algo:acc:glob_par_sect_end}
		\nonl \;  
		Constraints computation\;
	\end{multicols}
	\caption{aprod 1 with OpenMP and OpenACC\label{algo:aprod_1}}
\end{algorithm*}


\begin{algorithm*}
	\begin{multicols}{2}
		\nonl \textbf{\small aprod 2 with OpenMP}\;
		\SetNlSty{}{omp.}{}
		\textcolor{blue}{\texttt{\#pragma omp parallel private($pid$,$tid$,$nth$) shared($\vb*{N}$,$\vb*{x}$,$\mathbf{A}_\textbf{d}$,$\vb*{b}$)}} \;
		\{\;
		\tcp{ID number of the OpenMP thread} 
		\text{ }\text{ }\text{ }$tid$ = omp\_get\_thread\_num()\; \label{algo:omp:aprod2_tid}
		\tcp{Number of OpenMP threads}
		\text{ }\text{ }\text{ }$nth$ = omp\_get\_num\_threads();\; \label{algo:omp:aprod2_nth}
		\text{ }\text{ }\text{ }\For{$i\leftarrow \vb*{N}_\textbf{t}[tid][0]$ \KwTo $\vb*{N}_\textbf{t}[tid][1]$}  { \label{algo:omp:aprod2_Ntid}
			\nonl \;
			\tcp{Astrometric sect.}
			$k = i \times N_{\rm par}$ \;
			\For{$j\leftarrow 0$ \KwTo $N_{\rm Astro}$}{
				\nonl \;
				$\vb*{x}[j+ {\rm offset}[i]] = \vb*{x}[j + {\rm offset}[i]] + \mathbf{A}_\textbf{d}[k]\vb*{b}[i]$\;
				$k$++\;
			}
			\nonl \;
			\tcp{Attitude sect.}
			$k = i \times N_{\rm par} + N_{\rm Astro}$ \;
			\For{$j_1\leftarrow 0$ \KwTo $N_{\rm Axes}$}{
				$k_2 = j_1 \times N_{\rm DFA} + {\rm offset}[i] $\;
				\For{$j_2 \leftarrow 0$ \KwTo $N_{\rm ParAxis}$}{
					$\vb*{x}[j_2 + k_2] = \vb*{x}[j_2 + k_2] + \mathbf{A}_\textbf{d}[k]\vb*{b}[i]$\;
					$k$++\;
				}  
			}
			\nonl \;
			\tcp{Instrumental sect.}	        
			$k = i \times N_{\rm par} + N_{\rm Astro} + N_{\rm Att}$ \;
			\For{$j\leftarrow 0$ \KwTo $N_{\rm Instr}$}{
				$\vb*{x}[\mathcal{F}(i,j)$+${\rm offset}]$ = $\vb*{x}[\mathcal{F}(i,j)$+${\rm offset}]$+$\mathbf{A}_\textbf{d}[k]\vb*{b}[i]$\;
				$k$++\;
			}
			\nonl \;
			\tcp{Global sect.}
			$k = i \times N_{\rm par} + N_{\rm Astro} + N_{\rm Att} + N_{\rm Instr}$ \;
			\For{$j\leftarrow 0$ \KwTo $N_{\rm Glob}$}{
				$\vb*{x}[j+ {\rm offset}] = \vb*{x}[j + {\rm offset}] + \mathbf{A}_\textbf{d}[k]\vb*{b}[i]$\;
				$k$++\;
			}
		}
		\}\;
		Constraints computation\;


		
		\setcounter{AlgoLine}{0}
		\SetNlSty{}{acc.}{}
		\textbf{\small aprod 2 with OpenACC}\;
		\textcolor{red}{\texttt{\#pragma acc parallel}} \; \label{algo:acc:aprod_2_parallel_start} 
		\{\;
		\nonl \;
		\nonl \;
		\text{ }\text{ }\text{ }\textcolor{red}{\texttt{\#pragma acc loop}} \; \label{algo:acc:aprod_2_ext_loop}
		\text{ }\text{ }\text{ }\For{$i\leftarrow 0$ \KwTo $\vb*{N}[pid]$}  {
			\nonl \;
			\tcp{Astrometric sect.}
			$k = i \times N_{\rm par}$ \;
			\For{$j\leftarrow 0$ \KwTo $N_{\rm Astro}$}{
				\textcolor{red}{\texttt{\#pragma acc atomic}}\;
				$\vb*{x}[j+ {\rm offset}[i]] = \vb*{x}[j + {\rm offset}[i]] + \mathbf{A}_\textbf{d}[k+j]\vb*{b}[i]$\; \label{algo:acc:xpluseqAb_Astro}
			}
			\nonl \;
			\tcp{Attitude sect.}
			$k = i \times N_{\rm par} + N_{\rm Astro}$ \;
			\For{$j_1\leftarrow 0$ \KwTo $N_{\rm Axes}$}{
				$k_1 = j_1 \times N_{\rm ParAxis} $\;
				$k_2 = j_1 \times N_{\rm DFA} + {\rm offset}[i] $\;
				\For{$j_2 \leftarrow 0$ \KwTo $N_{\rm ParAxis}$}{
					\textcolor{red}{\texttt{\#pragma acc atomic}}\;
					$\vb*{x}[j_2 + k_2] = \vb*{x}[j_2 + k_2] + \mathbf{A}_\textbf{d}[k + j_2 + k_1]\vb*{b}[i]$\; \label{algo:acc:xpluseqAb_Att}
				}  
			}
			\nonl \;   	
			\tcp{Instrumental sect.}        
			$k = i \times N_{\rm par} + N_{\rm Astro} + N_{\rm Att}$ \;
			\For{$j\leftarrow 0$ \KwTo $N_{\rm Instr}$}{
				\textcolor{red}{\texttt{\#pragma acc atomic}}\;
				$\vb*{x}[\mathcal{F}(i,j) + {\rm offset}] = \vb*{x}[\mathcal{F}(i,j) + {\rm offset}] + \mathbf{A}_\textbf{d}[k+j]\vb*{b}[i]$\; \label{algo:acc:xpluseqAb_Instr}
			}
			\nonl \;
			\tcp{Global sect.}
			$k = i \times N_{\rm par} + N_{\rm Astro} + N_{\rm Att} + N_{\rm Instr}$ \;
			\For{$j\leftarrow 0$ \KwTo $N_{\rm Glob}$}{
				\textcolor{red}{\texttt{\#pragma acc atomic}}\;
				$\vb*{x}[j+ {\rm offset}] = \vb*{x}[j + {\rm offset}] + \mathbf{A}_\textbf{d}[k+j]\vb*{b}[i]$\; \label{algo:acc:xpluseqAb_Glob}
				\nonl \; 
			}
		}
		\} \; \label{algo:acc:aprod_2_parallel_end}
		\nonl \;
		Constraints computation\;
	\end{multicols}
	\caption{aprod 2 with OpenMP and OpenACC\label{algo:aprod_2}}
\end{algorithm*}

\section{The Marconi100 cluster and the production of the AVU--GSR code}
\label{sec:Production}

The application has been in production since 2014 under an agreement between Istituto Nazionale di Astrofisica (INAF) and CINECA, with the support of the Italian Space Agency (ASI). The code has run on all the Tier0 systems of CINECA and it is now employed by the Coordination Unit 3 (CU3) of DPAC for the AVU-GSR tasks. The entire process of AVU-GSR is managed by the Data Processing Center of Turin (DPCT) that is supervised by the Aerospace Logistics Technology Engineering Company (ALTEC) in collaboration with the Astrophysics Observatory of Turin (INAF-OATO).

The application is currently running on the CINECA supercomputer M100 which has 980 compute nodes with the following features:
\begin{enumerate}
	\item 2 sockets of 16 physical cores each, of the type IBM POWER9 AC922, with a processor speed of 3.1 GHz. Each physical core corresponds to 4 virtual cores, with a total of 128 ($2 \times 16 \times 4$) virtual cores per node;
	\item 4 GPUs of the type NVIDIA Volta V100, with a memory of 16 GB each, connected with Nvlink 2.0;
	\item 256 GB of RAM.
\end{enumerate}

So far, we have worked with a subset of the total number of observations. One of the most recent runs, which is representative of a typical execution, dealt with $\sim$8.4$\times10^6$ stars observed multiple times, with a total number of observations equal to $\sim$1.8$\times10^9$. The execution converged after $\sim$141000 iterations, with a time per iteration equal to $\sim$4.23 s. This results in a total elapsed time of $t_{\rm e} \simeq 166$ hours. The run was executed on 2 nodes of M100, on a total of $n_{\rm proc} = 32$ MPI processes (16 per node). For each MPI process, the code ran on $n_{\rm th} = 2$ OpenMP threads. The run occupied a memory of $\sim$10.6 GB per MPI process, correspondent to a total memory of $\sim$340 GB. With these features, the total computational time corresponds to $t_{\rm c} = t_{\rm e} \times n_{\rm proc} \times n_{\rm th} = 10624$ hours. 
The astrometric unknowns were retrieved with a micro-arcsecond resolution~\citep{Vecchiato_2018}.


\section{From the CPU to the GPU: OpenACC}
\label{sec:GPU}

To further improve the performance of the application, we ported it to the GPU by replacing OpenMP with OpenACC \citep[see][for the description of a semi-automatic methodology to parallelize scientific applications with the OpenACC parallelization model]{Aldinucci_2021}.

\subsection{Multi--GPU computation}
\label{sec:GPU_multi_GPU}

{The code runs on multiple GPUs, depending on the number of MPI processes. The MPI processes in each node are assigned to the GPUs of the node in a round-robin fashion, according to the command at line~\ref{algo:GPU_number_setting} of Algorithm~\ref{algo:Full_application}.}

\begin{figure*}[hbt!]
	\centering
	\includegraphics[width=0.75\textwidth]{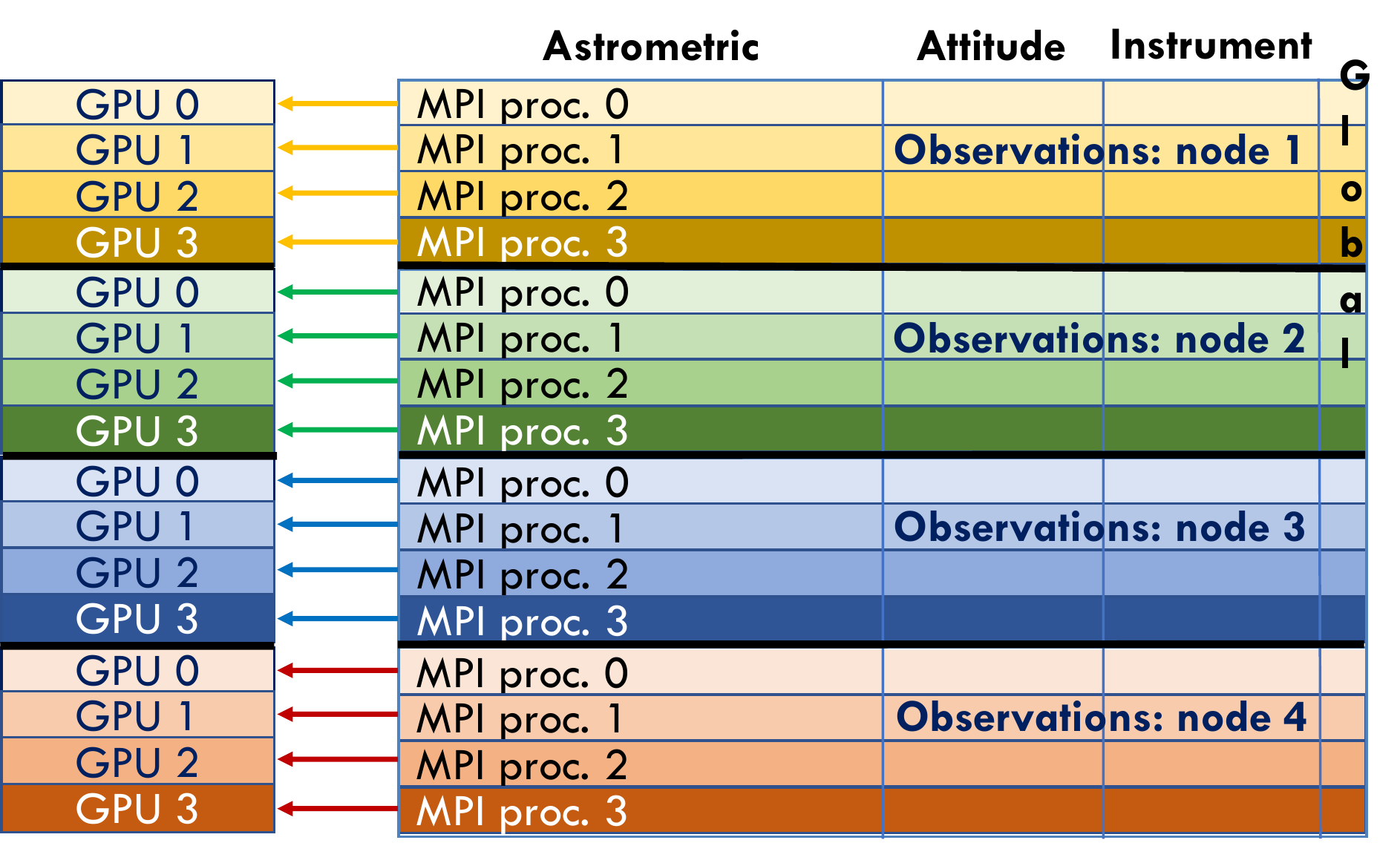}
	\caption{Parallelization scheme of the coefficient matrix of the system on four nodes of a computer cluster with four MPI processes and four GPUs per node. The {right part} of the figure represents the coefficient matrix and the {left part} shows the GPUs on the four nodes of the cluster. Each large stripe (yellow, green, blue, and brown) represents one node and the four colour-coded sub-stripes in each node correspond to the four MPI processes in the node, associated with the different GPUs of the node.}
	\label{fig:Par_Data_scheme_multi_GPU}
\end{figure*}

Figure~\ref{fig:Par_Data_scheme_multi_GPU} represents the parallelization scheme of the coefficient matrix on four nodes of a computer cluster with four MPI processes and four GPUs per node (e.g. M100).
The figure shows that the computation related to each MPI process is assigned to a different GPU. Running on a number of MPI processes per node equal to the number of GPUs per node, like in this example, represents the optimal configuration, as better explained in Section~\ref{sec:Performance_tests_Strong_scaling}.

\subsection{Data transfers}
\label{sec:GPU_data_transfers}

The GPU has $\sim$10$^2$ times more cores than the CPU and it is ideal to parallelize the computation of a large amount of data. When dealing with GPU programming, it is important to proper manage the data transfers from the host (CPU) to the device (GPU) (H2D) and vice versa (D2H). Indeed, if a code parallelized on the GPU becomes \textit{data-transfer bound}, namely the data transfers dominate the computation, its performance could worsen compared to the same code parallelized on the CPU. Managing the data copies is particularly important in iterative constructs, where it is essential to move as much data as possible at the beginning and/or at the end of the entire iteration cycle, and to reduce the copies at every step of the algorithm.

The data movements between the CPU and the GPU are regulated by particular directives placed in strategical points of the code. Specifically, we employed the following directives (see lines~\ref{algo:Full_application:copyin_beg_LSQR},~\ref{algo:Full_application:copyin_x_b_aprod1},~\ref{algo:Full_application:copyout_x_b_aprod1},~\ref{algo:Full_application:copyin_x_b_aprod2}, and~\ref{algo:Full_application:copyout_x_b_aprod2} of Algorithm~\ref{algo:Full_application}):
\begin{enumerate}
	\item \texttt{\#pragma acc enter data copyin()}, that copies H2D the arrays listed within the round brackets;
	\item \texttt{\#pragma acc exit data copyout()}, that copies D2H the arrays listed within the round brackets.
\end{enumerate}
In our port, we transfer $\sim$95\% of the data H2D at the beginning of the LSQR iterative cycle.
This is possible since these quantities are not modified during the computation on the GPU. Particularly, among these quantities there are (see line~\ref{algo:Full_application:copyin_beg_LSQR} of Algorithm~\ref{algo:Full_application}):
\begin{enumerate}
	\item $\mathbf{A}_\text{\bf d}$, the 1D double-precision array containing only the {non-zero elements} of the coefficient matrix $\mathbf{A}$ (the dense coefficient matrix);
	\item $\vb*{M}_\text{\bf i}$, the 1D single-precision array containing the number of the star and the index of the first attitude element for each observation of the original matrix $\mathbf{A}$;
	\item $\vb*{I}_\text{\bf c}$, the 1D single-precision array with the indexes that the instrumental elements of the dense matrix $\mathbf{A}_\text{\bf d}$ had in the original matrix $\mathbf{A}$;
	\item $\vb*{N}$, the 1D single-precision array containing the number of observations assigned to each MPI process.
\end{enumerate}
There is no reason to copy these data back to the CPU at the end of the entire cycle of iterations. For a typical execution as the one presented in Section~\ref{sec:Production}, that occupies a total memory of $\sim$340~GB, the quantities $\mathbf{A}_\text{\bf d}$, $\vb*{M}_\text{\bf i}$, and $\vb*{I}_\text{\bf c}$ occupy a memory of $\sim$246~GB, $\sim$27~GB, and $\sim$41~GB, respectively, whereas the amount of memory occupied by the array $\vb*{N}$ is negligible compared to the total.

The only arrays that need to be copied both H2D and D2H at every step of the LSQR algorithm are $\vb*{x}$, containing the unknowns of the problem, and $\vb*{b}$ containing the known terms of the system of equations (see lines~\ref{algo:Full_application:copyin_x_b_aprod1},~\ref{algo:Full_application:copyout_x_b_aprod1},~\ref{algo:Full_application:copyin_x_b_aprod2}, and~\ref{algo:Full_application:copyout_x_b_aprod2} of Algorithm~\ref{algo:Full_application}), since they are both updated at each iteration. 
The H2D and the D2H copies of $\vb*{x}$ and $\vb*{b}$ are performed both for the aprod 1 and for the aprod 2 functions but this does not imply a substantial slowdown of the code, since these vectors only represent the $\sim$5\% of the total memory. Indeed, for the execution presented above, the $\vb*{x}$ and $\vb*{b}$ arrays occupy a memory of $\sim$$0.23$~GB and $\sim$14~GB, respectively.

Managing the data copies in this way, that is moving the $\sim$95\% of the data H2D before the LSQR cycle and only the $\sim$5\% of the data H2D and D2H at each step of the LSQR, the time employed in memory transfers is much less than the time employed in calculations, namely the application is \textit{compute bound} rather than \textit{data-transfer bound}. This point is better addressed in Section~\ref{sec:Performance_tests_Speedup}. 

\begin{algorithm*}
	Import data from files\;
	Preconditioning\;
	\texttt{acc\_set\_device\_num(pid\%acc\_get\_num\_devices(acc\_device\_default),acc\_device\_default)}\; \label{algo:GPU_number_setting}
	\textcolor{red}{\texttt{\#pragma acc enter data copyin($\mathbf{A}_\textbf{d}$,$\vb*{M}_\textbf{i}$,$\vb*{I}_\textbf{c}$,$\vb*{N}$)}}\; \label{algo:Full_application:copyin_beg_LSQR}
	\textcolor{red}{\texttt{\#pragma acc enter data copyin($\vb*{x}$,$\vb*{b}$)}}\; \label{algo:Full_application:copyin_x_b_aprod2_IS}
	aprod 2 call (calculation of the initial solution)\;
	\textcolor{red}{\texttt{\#pragma acc exit data copyout($\vb*{x}$,$\vb*{b}$)}}\; \label{algo:Full_application:copyout_x_b_aprod2_IS}
	MPI reduce\; \label{algo:Full_application:MPI_reduce_1}
	\tcp{LSQR algorithm}
	\While{(conv. cond. $\vert$ $\vert$ max itn. reached)}{ 
		\textcolor{red}{\texttt{\#pragma acc enter data copyin($\vb*{x}$,$\vb*{b}$)}}\; \label{algo:Full_application:copyin_x_b_aprod1}
		aprod 1 call\;
		\textcolor{red}{\texttt{\#pragma acc exit data copyout($\vb*{x}$,$\vb*{b}$)}}\; \label{algo:Full_application:copyout_x_b_aprod1}
		MPI reduce\; \label{algo:Full_application:MPI_reduce_2}
		\textcolor{red}{\texttt{\#pragma acc enter data copyin($\vb*{x}$,$\vb*{b}$)}}\; \label{algo:Full_application:copyin_x_b_aprod2}
		aprod 2 call\;
		\textcolor{red}{\texttt{\#pragma acc exit data copyout($\vb*{x}$,$\vb*{b}$)}}\; \label{algo:Full_application:copyout_x_b_aprod2}
		MPI reduce\; \label{algo:Full_application:MPI_reduce_3}
		Variances and covariances computation\;
	}
   Print solution to files
	
	\caption{Structure of the entire Gaia AVU--GSR application in OpenACC \label{algo:Full_application}}
\end{algorithm*}

\subsection{Parallelization}
\label{sec:GPU_parallelization}

We defined the aprod modes 1 and 2 in two separate functions that we both ported to the GPU.
In aprod 1, we reorganized the structure of the code: instead of computing the astrometric, the attitude, the instrumental, and the global sections within a single external for loop iterating on the number of observations assigned to each MPI process (Algorithm~\ref{algo:aprod_1}, left {column}), we repeated the external loop four times, {once for every section} (Algorithm~\ref{algo:aprod_1}, right {column}). Then, we enclosed each of the newly arranged sections in a \texttt{\#pragma acc parallel private(sum)} directive, defining four parallel regions (lines~acc.\ref{algo:acc:astro_par_sect_beg}--\ref{algo:acc:astro_par_sect_end},~acc.\ref{algo:acc:att_par_sect_beg}--\ref{algo:acc:att_par_sect_end},~acc.\ref{algo:acc:instr_par_sect_beg}--\ref{algo:acc:instr_par_sect_end}, and~acc.\ref{algo:acc:glob_par_sect_beg}--\ref{algo:acc:glob_par_sect_end} of Algorithm~\ref{algo:aprod_1}).    
The \texttt{\#pragma} \texttt{acc} \texttt{parallel} directive starts a parallel execution on the current device and, to be effective, it requires an analysis by the programmer to ensure safe parallelism of the region of the code that is enclosed within the scope defined by the directive. The variable $sum$ is declared within the \texttt{private} clause (lines~acc.\ref{algo:acc:astro_par_sect_beg},~acc.\ref{algo:acc:att_par_sect_beg},~acc.\ref{algo:acc:instr_par_sect_beg}, and~acc.\ref{algo:acc:glob_par_sect_beg} of Algorithm~\ref{algo:aprod_1}), such that each GPU thread has a local copy of it.  
In each parallel region, we parallelized the most external for loop with the \texttt{\#pragma acc loop} directive (lines~acc.\ref{algo:acc:astro_par_loop},~acc.\ref{algo:acc:att_par_loop},~acc.\ref{algo:acc:instr_par_loop}, and~acc.\ref{algo:acc:glob_par_loop} of Algorithm~\ref{algo:aprod_1}). 

We inserted the entire aprod 2 in a \texttt{\#pragma} \texttt{acc} \texttt{parallel} region (Algorithm~\ref{algo:aprod_2}, lines~acc.\ref{algo:acc:aprod_2_parallel_start}--\ref{algo:acc:aprod_2_parallel_end}) and we parallelized the most external for loop with a \texttt{\#pragma acc loop} directive (Algorithm~\ref{algo:aprod_2}, line~acc.\ref{algo:acc:aprod_2_ext_loop}). With OpenACC, the variable \textit{tid}, that identifies the OpenMP thread in the CPU code, becomes unnecessary: therefore the for loop at line omp.\ref{algo:omp:aprod2_Ntid} of Algorithm~\ref{algo:aprod_2} does not iterate anymore from $\vb*{N}_\textbf{t}[tid][0]$ to $\vb*{N}_\textbf{t}[tid][1]$, but from 0 to $\vb*{N}[pid]$, where $pid$ is rank of the MPI process. On the lines of code that compute the operation $\vb*{x} = \vb*{x} + \mathbf{A} \times \vb*{b}$ in the astrometric, the attitude, the instrumental, and the global sections (lines acc.\ref{algo:acc:xpluseqAb_Astro}, acc.\ref{algo:acc:xpluseqAb_Att}, acc.\ref{algo:acc:xpluseqAb_Instr}, and acc.\ref{algo:acc:xpluseqAb_Glob} of Algorithm~\ref{algo:aprod_2}), we put a \texttt{\#pragma} \texttt{acc} \texttt{atomic} directive, preventing different GPU threads to simultaneously write the same elements of the array $\vb*{x}$.

\subsection{Compilation and optimization of the code}
\label{sec:GPU_compilation_optimization}

To compile the application, we employed the PGI compiler\footnote{\url{https://www.pgroup.com/resources/docs/19.1/x86/pgi-ref-guide/index.htm}}. Specifically, we prepared a Makefile and we compiled with the \texttt{-acc}, \texttt{-fast}, and \texttt{-ta=tesla:maxregcount:32} options, where:
\begin{enumerate}
	\item \texttt{-acc} enables the OpenACC directives; 
	\item \texttt{-fast} includes a set of flags to optimize the code;
	\item \texttt{maxregcount:n} specifies that the GPU employs a maximum number of registers equal to \texttt{n}. In particular, we set \texttt{n} to 32, since it optimizes the usage of the compute resources and the memory occupancy of a NVIDIA Volta V100 GPU, the device on which the code is tested, and it provides a speedup of the application. For better details, see Section~\ref{sec:Performance_tests_GPU_utilization}.
\end{enumerate}

The application uses the \textit{CFITSIO}\footnote{\url{https://heasarc.gsfc.nasa.gov/fitsio/}}, a library of C and Fortran subroutines for reading and writing data files in FITS data format. {Indeed, before the start of the LSQR iterations, the application converts the input data from FITS to binary format, read by the LSQR, and after the end of the LSQR iterations the solution is written to a binary file then converted to FITS format (Figure~\ref{fig:Flowchart_all}).}

\section{Performance tests}
\label{sec:Performance_tests}

We aim to build a GPU version of the Gaia AVU-GSR application that accelerates compared to the current MPI + OpenMP code. In this section, we compare the performance of the OpenMP and the OpenACC codes. To perform this task, we simulated a complete system of stars in the Milky Way, the attitude and the instrumental settings of the Gaia satellite, and the global parameter $\gamma$ of the PPN formalism, as described in Section~\ref{sec:Coeff_matrix}.

We ran the performance tests on the CINECA supercomputer M100, described in Section~\ref{sec:Production}. On M100, the memory of one node is of 256 GB and the memory of the 4 GPUs in each node is of $4 \times 16$ GB. For this reason, we cannot run the tests for the OpenACC code by simulating a system that occupies a memory larger than 64 GB per node. For safety reasons, to avoid memory overflows due to the GPU architecture, we decided to lower this limit to 40 GB per node. The OpenMP code is not subject to this limitation but we needed to set the same amount of memory in the two applications to compare their performance.

For a complete exploration of the performance of the two applications, we ran the tests both on a single node and on an increasing number of nodes, setting a system with a fixed amount of memory. We also ran the tests on more nodes, setting an amount of memory proportional to the number of nodes. 
Specifically, the three tests were run with the following configurations:

\begin{enumerate}
	\item \textit{Fixed memory, intra-node}: on 1 node of M100, on an increasing number of MPI processes, setting the parameters of the system such that it occupies a memory of 10 GB.\footnote{In this test, the memory is not set to 40 GB but to 10 GB because the codes are run on a number of MPI processes from 1 to 32. Since the OpenACC code executed on one MPI process runs on only one GPU, that has a memory of 16 GB, we cannot set a memory of 40 GB. It is logical to set the memory to 40 GB (the limit that we chose for 4 GPUs) divided by 4 (the number of GPUs in the node).};
	\item \textit{Fixed memory, inter-nodes}: on an increasing number of nodes of M100, up to 16 nodes, with 4 MPI processes per node and a system that occupies a memory of 40 GB;
	\item \textit{Proportional memory, inter-nodes}: on an increasing number of nodes of M100, up to 16 nodes, with 4 MPI processes per node, setting a system that occupies a memory of 40 GB per node (40 GB on one node, 80 GB on two nodes, and so forth). 
\end{enumerate}

{The tests are illustrated in Figure~\ref{fig:Scaling_tests}.} The code ran {up to convergence} of the LSQR algorithm, to increase the statistical meaning of the time measurements. On the vertical axis of all plots in Figure~\ref{fig:Scaling_tests}, we show the mean execution time of one LSQR iteration. The error on these measurements, represented as error bars, is provided by the standard deviation of the times of all the LSQR iterations. 

In the intra-node plot {(Figure~\ref{fig:Scaling_tests}a)}, the bottom horizontal axis represents the number of MPI processes, set to \{1,2,4,8,16,32\}. For the OpenMP version of the code, the top horizontal axis shows the number of OpenMP threads assigned to each MPI process. In particular, the number of OpenMP threads is set such that the product of the number of MPI processes and the number of OpenMP threads is equal to {32, the number of physical cores in one node} (Section~\ref{sec:Production}). Since we set the number of MPI processes  to \{1,2,4,8,16,32\}, the number of OpenMP threads is set to \{32,16,8,4,2,1\}. 

{To optimize each run in the intra-node and in the inter-nodes tests, we set the number of MPI processes per socket (\texttt{--ntasks-per-socket}) to the number of MPI processes per node (\texttt{--ntasks-per-node}) divided by two, except when we only consider one MPI process per node, since M100 is a two-socket platform.} As specified in Section~\ref{sec:Production}, each physical CPU core of M100 corresponds to 4 virtual cores. We required that each MPI process on each node was allocated on a different physical CPU core of the node since, as experienced on POWER9 architecture and suggested by the CINECA support for our specific case, allocating more MPI processes on different virtual cores of the same physical CPU causes a slowdown of the code. {To obtain this configuration, we set the number of virtual cores to allocate for each MPI process (\texttt{--cpus-per-task}) such that the product between \texttt{--ntasks-per-node} and \texttt{--cpus-per-task} is equal to 128, the total number of virtual cores per node of M100.} As detailed in the previous paragraph, to exploit all the physical CPUs in each node, we set the number of OpenMP threads (\texttt{OMP\_NUM\_THREADS}) such that the product of \texttt{--ntasks-per-node} and \texttt{OMP\_NUM} \texttt{\_THREADS} is equal to 32.  


Both the OpenMP and the OpenACC codes are launched with a \texttt{--map-by socket:} \texttt{PE=n} specification. This map defines the number of physical cores assigned to each MPI process in one socket and is set to the total number of physical cores present in a socket (16 on M100) divided by the number of MPI processes per socket assigned for the run. Setting this variable prevents each MPI process to run on more virtual cores belonging to the same physical core in each socket. This map is fundamental for the OpenMP code, that runs on the CPU, but it is basically irrelevant for the OpenACC code, that mainly runs on the GPU.

In the inter-nodes plots (Figures~\ref{fig:Scaling_tests}b and~\ref{fig:Scaling_tests}c), the bottom horizontal axis represents, instead, the number of nodes on which the codes run. Since the two inter-nodes tests run on 4 MPI processes per node, the OpenMP code is always parallelized on 8 OpenMP threads per MPI process. 
We ran the two inter-nodes test on \{1,2,4,8,16\} nodes, and thus on \{4,8,16,32,64\} total MPI processes.


\subsection{Fixed memory}
\label{sec:Performance_tests_Strong_scaling}

{Figures~\ref{fig:Scaling_tests}a and~\ref{fig:Scaling_tests}b} show the intra-node and the inter-nodes performance tests of the code, respectively, with a fixed memory configuration. Both plots demonstrate that the OpenACC code is more performant than the OpenMP code almost along the entire range of considered computational resources.

In the intra-node case, the gain in performance over the OpenMP version increases when the OpenACC code runs on a number of MPI processes larger or equal to 4. This is due to the fact that, in these configurations, the code runs on the four GPUs of the node, whereas when the number of MPI processes is set to 1 and 2, the code only exploits 1 and 2 GPUs, respectively {(line~\ref{algo:GPU_number_setting} of Algorithm~\ref{algo:Full_application})}. 
When running on a number of MPI processes $\geq$ 4, the time of one iteration of the OpenACC code remains nearly constant. For this reason, we could say that the optimal configuration to run the OpenACC code is on 4 MPI processes per node, since we obtain the best performance employing the minor number of computing resources: in this setting, all the GPUs of the node are exploited and only one MPI process is assigned to each GPU. This is the configuration on which the GPU code will run when in production. 
	
Concerning the MPI + OpenMP code, we observe that the response time decreases when the MPI processes increase and the OpenMP threads decrease. Indeed, the OpenMP parallelization is only employed within the two aprod functions, whereas MPI parallelizes the entire structure of the code, combining among the MPI processes the results obtained from the two aprod functions at each step of the LSQR algorithm. In particular, we can observe from Figure~\ref{fig:Scaling_tests}a that the optimal configuration for the OpenMP code is to run on 16 MPI processes and 2 OpenMP threads. We note that this configuration was employed for the in-production run presented as an example in Section~\ref{sec:Production} {and for all the other production runs}. Computing the speedup of the OpenACC code over the OpenMP code as the ratio between the iteration times achieved in the two optimal settings, 4 MPI processes for the OpenACC code and 16 MPI processes + 2 OpenMP threads for the OpenMP code, we obtain a factor of $\eta = 1.20 \pm 0.02$. In general, from 4 MPI processes on, the ratio between the average times of one LSQR iteration of the OpenMP and the OpenACC codes is nearly constant, around $1.3$, and reaches a maximum of $1.5$ when comparing the two values on 32 MPI processes.  

In the inter-nodes case, the ratio between the OpenMP and the OpenACC average times remains nearly constant along the entire range of nodes on which the codes were run. This is explained by the fact that the two codes always run in the same configuration, on 4 MPI processes per node. Specifically, the OpenACC code always runs in its optimal setting. Specifically, the average ratio, with its dispersion, is $1.39 \pm 0.06$, consistent with the results obtained for the intra-node case. 

{We note that the first point of the plot in Figure~\ref{fig:Scaling_tests}b corresponds to a run of the two codes on 4 MPI processes on one node and has an ordinate, i.e. the time for one iteration, $\sim$4 times larger than the ordinate of the third point of the plot in Figure~\ref{fig:Scaling_tests}a, that runs in the same configuration.} This is explained by the fact that the inter-nodes test computes a system that occupies an amount of memory 4 times larger than the memory occupied by the system in the intra-node test, as specified in the numbered list in Section~\ref{sec:Performance_tests}.

\subsubsection{Strong scaling}
\label{sec:Strong_scaling}

We investigate the strong scaling of the OpenMP and the OpenACC applications across the nodes. However, porting the Gaia code to the GPU is not intended to improve its scalability compared to the CPU code but its performance, to obtain scientific results in more reasonable timescales. In fact, we do not expect the scalability of the OpenACC code to substantially change compared to the OpenMP code, since the two applications have the same structure.

The similar scaling behaviour of the OpenMP and the OpenACC codes is confirmed by {Figure~\ref{fig:Scaling_tests_speedup}a,} that corresponds to {Figure~\ref{fig:Scaling_tests}b}, where the iteration time is replaced by the speedup, computed as:
\begin{equation}
\label{eq:Speedup}
	S = \frac{t_1}{t_n}.
\end{equation}
In Eq.~\eqref{eq:Speedup}, $t_1$ is the time of one {average} iteration of the LSQR algorithm on one node and $t_n$ is the iteration time on an increasing number of nodes. The error bars are calculated by propagating the uncertainties on $t_1$ and $t_n$. {For comparison, we show as a black dashed line the ideal speedup relation}. 

For both codes, the perfect scaling is achieved only up to 2 nodes and for a larger number of nodes the two scalability curves depart from the one-to-one line. Specifically, on 16 nodes, the OpenMP and the OpenACC codes reach a maximum speedup of $9.91$ and $9.57$, respectively, which translates to a parallel efficiency of $9.91/16 = 0.62$ and $9.57/16 = 0.60$. This moderate scalability can be explained by the non-parallelizable parts of the two applications, such as atomic operations, and by the communications among the MPI processes scheduled across the nodes.

\subsection{Proportional memory}
\label{sec:Performance_tests_Weak_scaling}

\begin{figure*}[hbt!]
	\centering
	\includegraphics[width=0.8\textwidth]{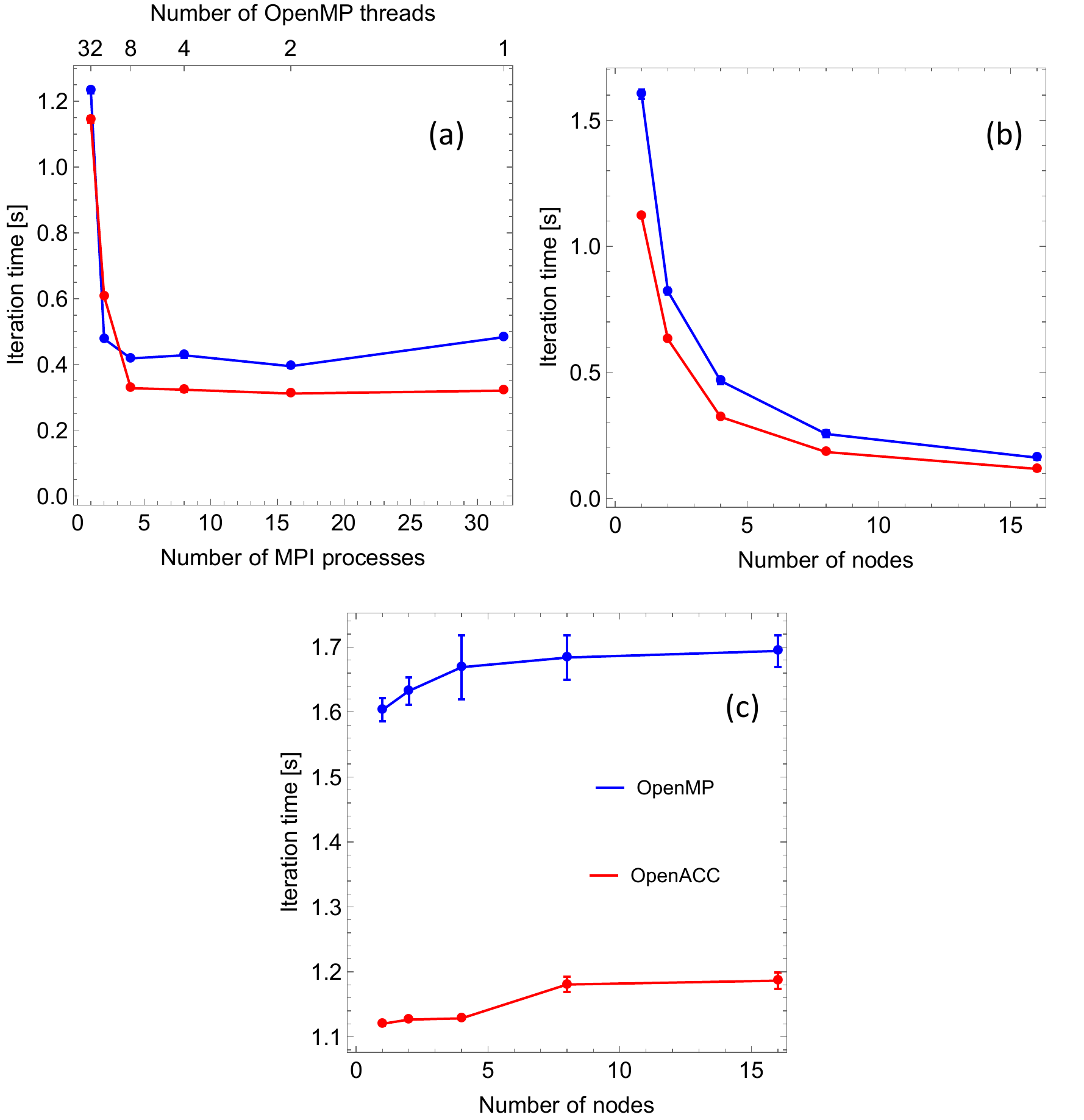}
	\caption{Performance tests for the OpenMP (blue lines and error bars) and the OpenACC (red lines and error bars) codes. The vertical axis represents the mean execution time of one LSQR iteration. {\textit{Figure~\ref{fig:Scaling_tests}a}:} intra-node fixed memory performance test. The bottom axis shows the number of MPI processes on which the codes run and the top axis the number of OpenMP threads assigned to each MPI process for the OpenMP code. 
	{\textit{Figure~\ref{fig:Scaling_tests}b}}: inter-nodes fixed memory performance test. The bottom axis shows the number of nodes on which the codes run. {\textit{Figure~\ref{fig:Scaling_tests}c}}: inter-nodes proportional memory performance test. The bottom axis is as in {Figure~\ref{fig:Scaling_tests}a}.}
	\label{fig:Scaling_tests}
\end{figure*}

{Figure~\ref{fig:Scaling_tests}c} shows the inter-nodes performance test of the codes where the system occupies a memory proportional to the amount of computational resources. Also in this case, the two codes always run on 4 MPI processes per node and the OpenACC code is always in its optimal configuration. {The mean gain of the OpenACC code over the OpenMP code, $1.44 \pm 0.02$, is in agreement with the results obtained in the two \textit{fixed memory} tests. }

\subsubsection{Weak scaling}
\label{sec:Weak_scaling}

\begin{figure*}[hbt!]
	\centering
	\includegraphics[width=0.9\textwidth]{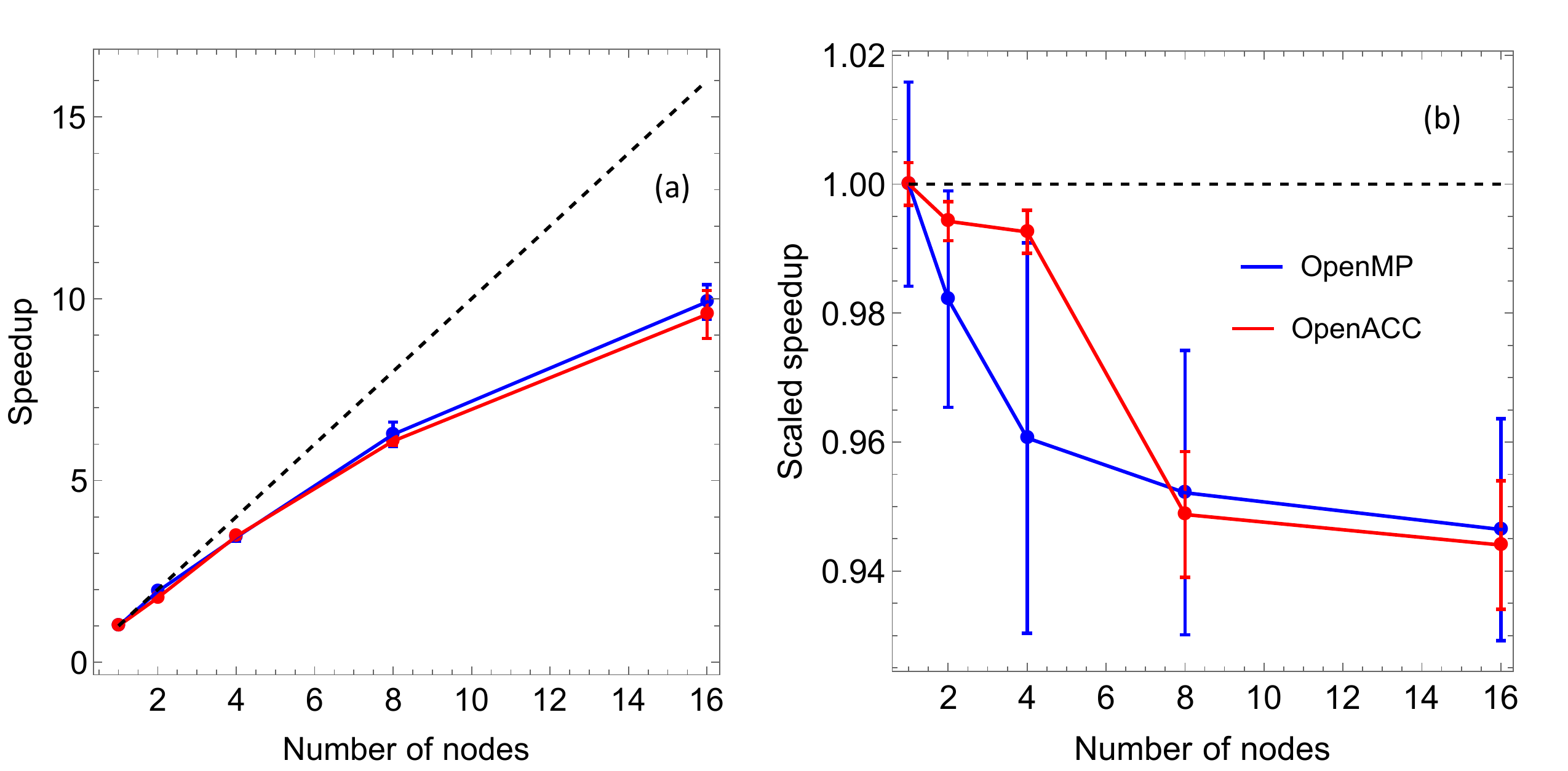}
	\caption{Fixed memory {(\textit{a})} and proportional memory {(\textit{b})} {inter-nodes} performance tests for the OpenMP (blue lines and error bars) and the OpenACC (red lines and error bars) codes. { \textit{Figure~\ref{fig:Scaling_tests_speedup}a}:} strong scaling of the OpenMP and the OpenACC codes. The black dashed line represents {the ideal speedup.}  
	{\textit{Figure~\ref{fig:Scaling_tests_speedup}b}:} weak scaling of the OpenMP and the OpenACC codes. The black dashed line represents the ideal Gustafson's law~\citep{Gustafson_1988}. {Figures~\ref{fig:Scaling_tests_speedup}a and ~\ref{fig:Scaling_tests_speedup}b correspond to Figures~\ref{fig:Scaling_tests}b and~\ref{fig:Scaling_tests}c, respectively.}}
	\label{fig:Scaling_tests_speedup}
\end{figure*}

We investigate the weak scaling of the OpenMP and of the OpenACC applications across the nodes. {Figure~\ref{fig:Scaling_tests_speedup}b} corresponds to {Figure~\ref{fig:Scaling_tests}c}, where the iteration time is replaced by the {scaled speedup}. The {scaled speedup} and its error are calculated as in Section~\ref{sec:Strong_scaling}. For comparison, we show as a black dashed line the $S = 1$ relation, where $S$ is the {scaled speedup} (Eq.~\ref{eq:Speedup}), representing the ideal trend of weak scalability, as stated by the Gustafson's law~\citep{Gustafson_1988}. The plot shows that the weak scalability curves of the OpenMP and the OpenACC codes are in agreement with each other and that the weak scaling is quite well satisfied for both applications, since the minimum {scaled speedup} is of $0.95$ for the OpenMP code and of $0.94$ for the OpenACC code. The mean iteration time passes from a minimum of $1.60$~s (OpenMP) and $1.12$~s (OpenACC) when the codes run on one node to a maximum of $1.69$~s (OpenMP) and $1.19$~s (OpenACC) when the codes run on 16 nodes.

\subsection{Detailed analysis of the speedup}
\label{sec:Performance_tests_Speedup}

We now investigate the origin of the speedup of the OpenACC code over the OpenMP code by comparing the elapsed times of the different regions of the two applications and by evaluating how much the memory transfers H2D/D2H of the OpenACC code affect its performance. Specifically, we compare a run of the OpenACC and of the OpenMP codes in their optimal configurations on one node of M100 (4 MPI processes for the OpenACC code and 16 MPI processes + 2 OpenMP threads for the OpenMP code). In this analysis, we do not consider the regions before the first call of the aprod 2 {function}, to calculate the initial solution, and of the region after the end of the LSQR iteration cycle (see Figure~\ref{fig:Flowchart_all}), since they are comparable between the two codes, given that they are not ported to the GPU in the OpenACC application. The simulated systems occupy a global memory of 10 GB, as in the intra-node performance test. 

\begin{figure*}[hbt!]
	\centering
	\includegraphics[width=1.0\textwidth]{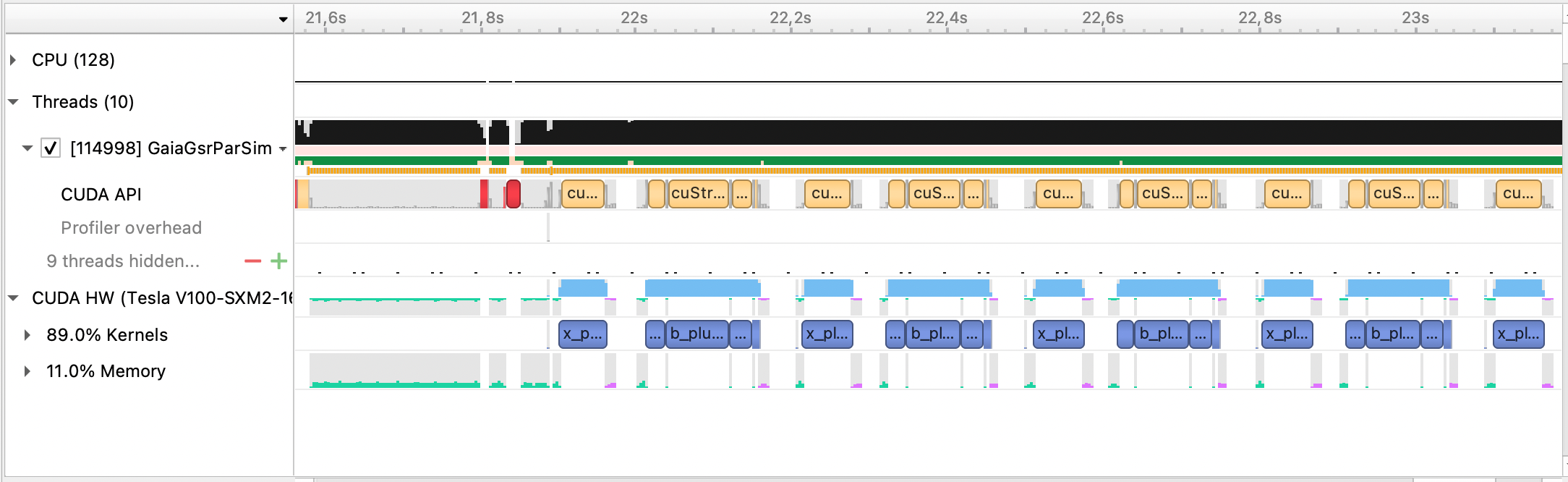}
	\caption{Output of the NVIDIA Nsight Systems profiler tool for a run with 4 iterations of the LSQR algorithm of the Gaia AVU-GSR application ported with OpenACC {to} 4 MPI processes (4 GPUs) of one node on M100. The system occupies a memory of 10 GB. The output shows the portion of the OpenACC code from line~\ref{algo:Full_application:copyin_beg_LSQR} to line~\ref{algo:Full_application:MPI_reduce_3} of Algorithm~\ref{algo:Full_application}. The blue areas, with the name of the kernel, represent the computation regions parallelized with an OpenACC directive (GPU kernels), the green and the purple areas represent the H2D and D2H memory transfers, respectively, and the white areas represent the regions still computed on the CPU.}
	\label{fig:Complete_profiler_output}
\end{figure*}

Figure~\ref{fig:Complete_profiler_output} shows the output of the NVIDIA Nsight Systems profiler tool\footnote{\url{https://developer.nvidia.com/nsight-systems}} for the run of the OpenACC code. We considered a run with 4 iterations of the LSQR algorithm. The NVIDIA Nsight Systems profiler is an analysis tool to visualize all the regions of a GPU-ported application: (I) computed by a GPU kernel (blue regions with the name of the correspondent kernel), (II) that involve memory transfers H2D (green regions) and D2H (purple regions), and (III) still computed on the CPU (white regions). This tool is particularly useful to suggest how to optimize an in-development application, like the Gaia AVU-GSR one. Specifically, Figure~\ref{fig:Complete_profiler_output} shows the relevant portion of the output of the OpenACC code, from line~\ref{algo:Full_application:copyin_beg_LSQR} to line~\ref{algo:Full_application:MPI_reduce_3} of Algorithm~\ref{algo:Full_application}, to be compared with the correspondent part of the OpenMP code.

\begin{figure*}[hbt!]
	\centering
	\includegraphics[width=1.0\textwidth]{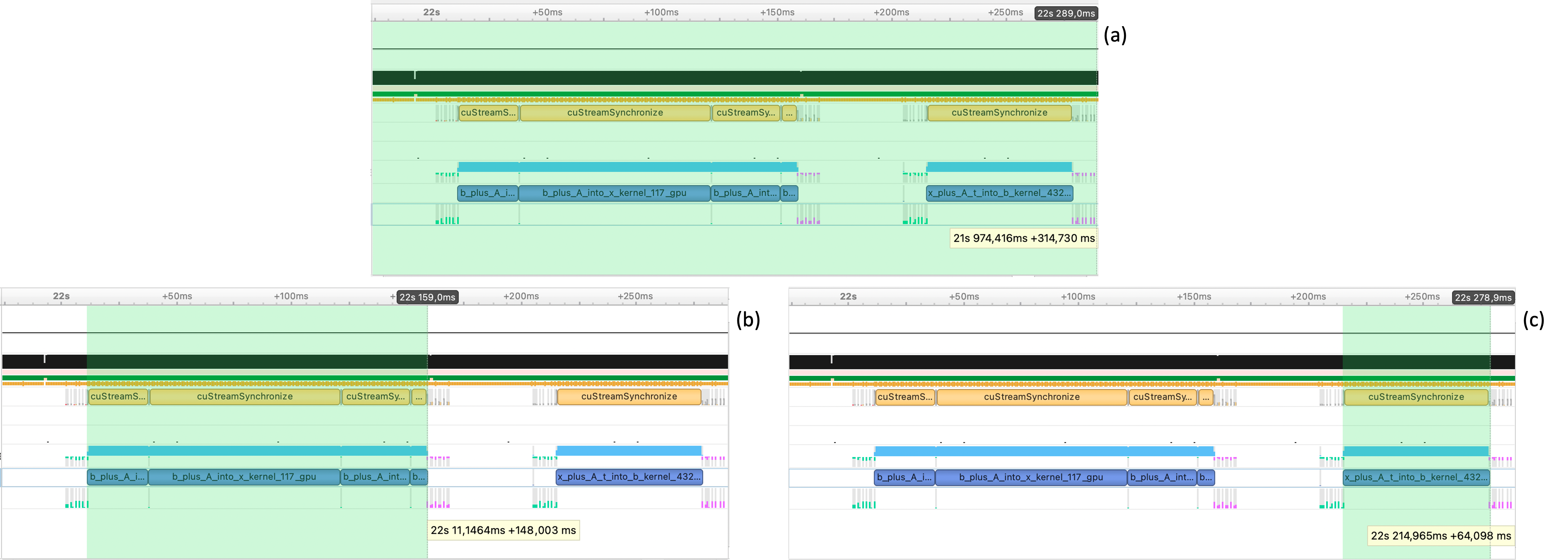}
	\caption{ {\textit{Figure~\ref{fig:Profiler_output_1iteration_aprod1_aprod2}a}}: Zoom-in of one iteration of the LSQR cycle of the run shown in Figure~\ref{fig:Complete_profiler_output}, with superimposed its elapsed time. {\textit{Figure~\ref{fig:Profiler_output_1iteration_aprod1_aprod2}b}}: Same as {Figure~\ref{fig:Profiler_output_1iteration_aprod1_aprod2}a} but with highlighted the elapsed time of the only aprod 1 region. { \textit{Figure~\ref{fig:Profiler_output_1iteration_aprod1_aprod2}c}}: Same as {Figure~\ref{fig:Profiler_output_1iteration_aprod1_aprod2}a} but with highlighted the elapsed time of the only aprod 2 region.}
	\label{fig:Profiler_output_1iteration_aprod1_aprod2}
\end{figure*}

The most expensive computation regions are the executions of the aprod 1 and 2 functions. {Figure~\ref{fig:Profiler_output_1iteration_aprod1_aprod2}a} shows a zoom-in of one iteration of the run of the OpenACC code, illustrated in Figure~\ref{fig:Complete_profiler_output}, where the aprod 1 and aprod 2 kernels, identified with the blue areas labelled as ``b\_plus...'' (aprod 1) and ``x\_plus...'' (aprod 2), are more visible. The times employed by these regions are directly measured from the profiler, for the OpenACC code (see {Figures~\ref{fig:Profiler_output_1iteration_aprod1_aprod2}b and~\ref{fig:Profiler_output_1iteration_aprod1_aprod2}c}, where these times are highlighted on the corresponding kernels), and with the \texttt{MPI\_Wtime()} function, for the OpenMP code. 

In the OpenACC code, the aprod 1 computation is divided into four kernels, one for the astrometric, one for the attitude, one for the instrumental, and one for the global part of the system of equations, as we can see in {Figure~\ref{fig:Profiler_output_1iteration_aprod1_aprod2}b} and in the right part of Algorithm~\ref{algo:aprod_1}. The total elapsed time of these four regions of the aprod 1 is of $t_{\rm a1,ACC} \sim 0.15$~s (see {Figure~\ref{fig:Profiler_output_1iteration_aprod1_aprod2}b}), and the correspondent time for the OpenMP code is of $t_{\rm a1,OMP} \sim 0.12$~s. This clearly means that the speedup of the OpenACC application over the OpenMP code is not due to the aprod 1 function but that, instead, the OpenACC code looses in performance compared to the OpenMP counterpart in executing this function. Specifically, the ratio between the OpenMP and OpenACC times of the aprod 1 is of $\sim$$0.8$.

From {Figure~\ref{fig:Profiler_output_1iteration_aprod1_aprod2}c} we can see that the aprod 2 kernel employs $t_{\rm a2,ACC} \sim 0.064$~s for the OpenACC code. Instead, in the OpenMP code, the aprod 2 function employs $t_{\rm a2,OMP} \sim 0.23$~s. In this case, the OpenACC code clearly accelerates compared to the OpenMP code, with a speedup of $\sim$$3.6$. Dividing the sum of the times of the aprod 1 and 2 regions for the OpenMP and the OpenACC codes, we obtain a speedup of $\sim$$1.6$, a bit larger than the one found in the previous sections. This is explained by the fact that in the OpenACC code we loose some time in copying the data H2D and D2H for every iteration, operation that is not performed in the OpenMP code. 

In Figure~\ref{fig:Profiler_output_1iteration_aprod1_aprod2} and in Algorithm~\ref{algo:Full_application}, we can see that we copy twice the $\vb*{x}$ and $\vb*{b}$ arrays both H2D and D2H at each iteration of the LSQR algorithm. These four copies employ a total time of  $t_{\rm Mem} \sim 0.04$~s, that is smaller than the computation times of both the aprod 1 and 2 regions. The white regions, namely all the operations still performed on the CPU that include minor I/O and the reduce operations among the MPI processes, employ a total time of $t_{\rm CPU} \sim 0.064$~s, comparable to the corresponding time in the OpenMP code. We estimate the total speedup as:

\begin{equation}
\label{eq:speedup}
	\eta' = \frac{t_{\rm a1,OMP} + t_{\rm a2,OMP} + t_{\rm CPU}}{t_{\rm a1,ACC} + t_{\rm a2,ACC} + t_{\rm Mem} + t_{\rm CPU}} \sim 1.3, 
\end{equation}
which is consistent with the value found in Section~\ref{sec:Performance_tests_Strong_scaling}. This speedup is mainly due to the acceleration of the computation of the aprod 2 region. 

We have seen that, for each iteration of the OpenACC code, the time involved in data copies is of $t_{\rm Mem} \sim 0.04$~s, whereas the time involved in kernel computation is of 

\[
   t_{\rm a1,ACC}+t_{\rm a2,ACC} \sim 0.15 + 0.064 \sim 0.21\textbf{ }\mathrm{s}.
\]

The data copies represent the $\sim$$18.7$\% of the time employed in kernel computation, which means that the code is \textit{compute bound} rather than \textit{data-transfer bound}. This is a consideration only for one iteration. Yet, if we observe the left part of Figure~\ref{fig:Complete_profiler_output}, we can see that for the entire run the $89.0$\% of the time is due to kernel computation and only the $11.0$\% of the time is due to memory transfers. The \textit{compute bound} consideration can thus be extended to the entire application.

\subsection{GPU utilization}
\label{sec:Performance_tests_GPU_utilization}

The OpenACC code runs on 32 GPU registers. This parameter is particularly important for a GPU code. On a NVIDIA V100 GPU, setting the number of registers to 32 might be a logical and optimal choice, since the NVIDIA V100 architecture is organized such that groups of 32 registers see the same cache memory and are subject to the same operation in a Single Instruction Multiple Data (SIMD)-like fashion. In the software, this is encoded in the size of a warp, a logical block of 32 threads that always perform the same operations simultaneously. Each warp is directly mapped on each block of 32 registers. 

\begin{figure*}[hbt!]
	\centering
	\includegraphics[width=\textwidth]{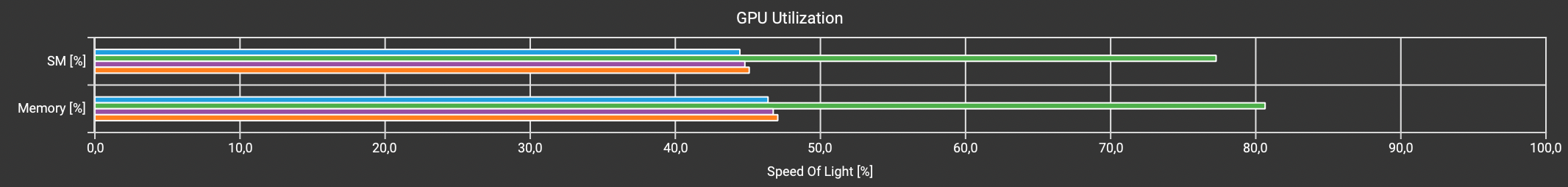}
	\caption{Percentage of utilization of the compute (SM) and of the memory resources of the GPU compared to the theoretical maximum (\textit{Speed Of Light} metric) when the OpenACC code is run on 32 (green bar), 64 (light blue bar), 128 (purple bar), and 42 (orange bar) GPU registers, as set during compilation. The plot is performed with the NVIDIA Nsight Compute profiler and refers to a system that occupies 10 GB of memory and runs on 1 GPU.}
	\label{fig:GPU_Utilization}
\end{figure*}

To verify whether 32 GPU registers actually correspond to the optimal configuration, we exploited the NVIDIA Nsight Compute profiler tool\footnote{\url{https://developer.nvidia.com/nsight-compute}}. With this profiler, we compared the \textit{Speed Of Light} metric, that calculates the percentage of utilization of the compute (SM) and of the memory resources of the GPU compared to the theoretical maximum, of four configurations, where we set the number of registers to 32, 64, 128, and 42 (green, light blue, purple, and orange bars in Figure~\ref{fig:GPU_Utilization}, respectively), three numbers multiples of 32 and one number that is not a multiple of 32. Figure~\ref{fig:GPU_Utilization} refers to a system that occupies 10 GB of memory and that runs on 1 GPU.

Whereas with 32 registers $\sim$80\% of the available compute and memory performance of the GPU are utilized, for the other three cases these reduce to $\sim$45\% and are comparable to each other. For 32 registers, the resources of the device are better exploited. A higher \textit{Speed Of Light} metric corresponds to a better performance of the code: whereas the mean iteration times for the 64-, 128-, and 42-registers cases are of $1.30$~s, $1.27$~s, and $1.25$~s, the mean iteration time for the 32-registers case is of $1.05$~s, which implies a speedup of $1.24$, $1.21$, and $1.19$ compared to the other three configurations.

\section{Conclusions and Future Work}
\label{sec:Conclusions_and_future_works}

{ 

We ported to the GPU with OpenACC the Gaia AVU--GSR solver, that aims to find the positions and the proper motions of $\sim$10$^8$ stars in our galaxy, besides the attitude and the instrumental specifications of the Gaia satellite and the parameter $\gamma$ of the PPN formalism. The application, originally parallelized on the CPU with MPI+OpenMP, solves, with the iterative LSQR algorithm, a system of linear equations, where the coefficient matrix is large and sparse.



The main target of this analysis is to explore the feasibility of porting this application to a GPU environment through a preliminary work based on the OpenACC library. This study, along with the investigation of the performance improvement, is propaedeutic to the final goal of a CUDA port and to a better optimization of the algorithm, to exploit at most the GPU architecture. 

To perform this preliminary port, we replaced the OpenMP part with the OpenACC language. In the OpenACC port, we moved the $\sim$95\% of the data H2D before the start of the LSQR cycle to limit the copies per iteration, which makes the code \textit{compute bound} rather than \textit{data-transfer bound}.


We compared the performance of the OpenMP and the OpenACC applications on M100 by running systems that occupy a memory both constant and proportional to the amount of computing resources.
The OpenACC code presents a speedup of $\sim$1.2--1.5 over the OpenMP code, and its optimal configuration is obtained by running on 4 MPI processes per node, which allows to exploit all the GPUs of the node assigning a single MPI process per GPU. With a speedup of $\sim$1.3, a typical execution of the AVU--GSR solver, as the one presented in Section~\ref{sec:Production}, passes from an elapsed time of $\sim$166 hours to $\sim$128 hours, saving the $\sim$23\% of the total time, in agreement with the estimate presented in Section~\ref{sec:Related_works}. This speedup is mainly driven by the port of the aprod 2 function to the GPU, which accelerates $\sim$3.6 times over the CPU version. 
We point out that, to control the GPUs, the MPI processes are the most logical choice compared to the OpenMP threads. A OpenMP + OpenACC version of this code would follow a completely different structure compared to the MPI + OpenMP application and it would be limited to run on only one node, which is not possible for systems with a size even much smaller than the expected final data set of Gaia.
	
The \textit{proportional memory} test shows that both the OpenMP and the OpenACC applications satisfy the weak scalability, since the average time of one LSQR iteration maintains nearly constant with the number of nodes, proportional to the memory occupied by the system. 

To exploit at best the compute and the memory resources of the GPU, we compiled the OpenACC code with 32 GPU registers, which entail the optimal performance compared to a different number of registers.

}

Additional {analyses} to further accelerate the code are already underway or planned. A possible way to explore is the asynchronous computation of the {CPU and the GPU code regions.} 
Some tests have demonstrated that, with the current structure of the code, we do not obtain a significant gain in performance if we asynchronously run, for example, {the aprod 1 and 2 regions, ported to the GPU with OpenACC, and their respective constraints sections, run on the CPU.} This is due to the fact that, although running on the CPU, {the two constraints regions} are $\sim$100-1000 faster than the aprod 1 and 2 functions. For the same reason, also a port of these two regions to the GPU would not significantly improve the code speed.

Another possibility would be {to reduce the copies} of the  $\vb*{x}$ and $\vb*{b}$ vectors at every iteration. However, the main future aim for this application is to port it to the GPU by replacing OpenACC with CUDA and by further optimizing the algorithm. {CUDA is a low-level parallelization language} that would imply a reorganization of some parts of the code to better match the GPU architecture, which would entail a more efficient parallelization and, thus, a higher speedup. This CUDA port is already in progress and the preliminary results are very optimistic. The acceleration factor over the OpenMP application might be around 10. With the CUDA port, we might also explore the advantages of the asynchronous {computation} of different GPU regions and of CPU and GPU regions.
	
We began to define the CUDA code by following the architecture of the V100 GPUs on M100 and we plan to readjust it to fit the architecture of the A100 GPUs that will be present on the pre-exascale system Leonardo, a supercomputer of CINECA that will be operational {at the end of} 2022. Given that the memory and the number of streaming multiprocessors of the A100 GPUs of Leonardo {will be larger than on the V100 GPUs} of M100, we would expect a further speedup of the code when it will run on this system. This is the object of a paper in preparation.
	
To conclude, the preliminary tests presented in this work provide essential information about the potential performance and scaling properties of the GPU-ported Gaia AVU-GSR code in perspective of exascale systems. These properties could be extrapolated to a class of codes that analogously solve an inverse problem for a large-sized system, that, as we have seen in Section~\ref{sec:Related_works}, we can retrieve in several contexts. However, these tests have to be extended to provide more exhaustive information. The largest system that we computed in the tests presented in this paper occupies a memory of 640 GB (40 GB per node on 16 nodes) and contains $\sim$$3 \times 10^9$ observations of the Milky Way stars, $\sim$2 orders of magnitude less than in the expected final data set of the Gaia mission, i.e.  $\sim$$10^{11}$ observations. A system with $\sim$$10^{11}$ observations will occupy a memory of $\sim$$\frac{10^{11}}{3 \times 10^9} \times 640 = 21333$~GB, which will require 356 nodes of M100 to be solved, by running with 60~GB per node, nearly the maximum allowed for a GPU code. To better explore the behaviour of this code and of other LSQR-based applications on exascale systems, we aim to extend the presented tests up to a larger amount of computing resources, for systems with an increasing size up to realistic science cases. This is what we plan to do with the CUDA code both on M100 and on the pre-exascale cluster Leonardo.

\section*{CRediT authorship contribution statement}
\label{sec:CRediT}
\textbf{Valentina Cesare:} Software, Validation, Formal analysis, Investigation, Writing - Original Draft, Visualization. \textbf{Ugo Becciani:} Term, Conceptualization, Software, Validation, Writing - Review \& Editing, Supervision, Project administration. \textbf{Alberto Vecchiato:} Software, Writing - Review \& Editing, Supervision. \textbf{Mario Gilberto Lattanzi:} Writing - Review \& Editing, Supervision, Funding acquisition. \textbf{Fabio Pitari:} Software, Investigation, Writing - Review \& Editing. \textbf{Mario Raciti:} Software, Investigation, Writing - Review \& Editing. \textbf{Giuseppe Tudisco:} Software, Investigation, Writing - Review \& Editing. \textbf{Marco Aldinucci:} Methodology, Writing - Review \& Editing. \textbf{Beatrice Bucciarelli:} Writing - Review \& Editing, Supervision.



\section*{Declaration of competing interest}
The authors declare the following financial interests/ personal relationships which may be considered as potential competing interests: Mario Gilberto Lattanzi has patent Gaia AVU–GSR parallel solver licensed to Mario Gilberto Lattanzi, Alberto Vecchiato, Beatrice Bucciarelli, Roberto Morbidelli, Ugo Becciani, Valentina Cesare. Alberto Vecchiato has patent Gaia AVU–GSR parallel solver licensed to Mario Gilberto Lattanzi, Alberto Vecchiato, Beatrice Bucciarelli, Roberto Morbidelli, Ugo Becciani, Valentina Cesare. Beatrice Bucciarelli has patent Gaia AVU–GSR parallel solver licensed to Mario Gilberto Lattanzi, Alberto Vecchiato, Beatrice Bucciarelli, Roberto Morbidelli, Ugo Becciani, Valentina Cesare. Ugo Becciani has patent Gaia AVU–GSR parallel solver licensed to Mario Gilberto Lattanzi, Alberto Vecchiato, Beatrice Bucciarelli, Roberto Morbidelli, Ugo Becciani, Valentina Cesare. Valentina Cesare has patent Gaia AVU–GSR parallel solver licensed to Mario Gilberto Lattanzi, Alberto Vecchiato, Beatrice Bucciarelli, Roberto Morbidelli, Ugo Becciani, Valentina Cesare.

\section*{Data availability}
The data that has been used is confidential.

\section*{Acknowledgements}
\label{sec:Acknowledgements}
	
	{We sincerely thank the referee, whose comments largely improved and clarified the presentation of our results.} We sincerely thank Dr. Aswin Kumar of NVIDIA, one of the mentors of the CINECA GPU Hackathon Digital Event of 2021, for the precious indications provided during the event that allowed to achieve the current speedup of the OpenACC code {over} the OpenMP code and for his help with the usage of the NVIDIA profilers. We also thank Dr. Massimiliano Guarrasi of CINECA, for the time that he dedicated to explain the basis of GPU architecture, and the organizers of the CINECA course ``Programming paradigms for GPU devices'', held on 9$^{\rm th}-11^{\rm th}$ June 2021, for their availability to deepen the fundamental aspects of GPU programming that were exploited to parallelize our application in a more efficient way.

\section*{Funding}
\label{sec:Funding}

This work was supported by the Italian Space Agency (ASI) [Grant Number: 2018-24-HH.0], in support of the Italian participation to the Gaia mission. This work was also supported by Consorzio Interuniversitario Nazionale per l'Informatica, under the project EUPEX, EC H2020 RIA, EuroHPC-02-2020 [Grant Agreement: 101033975].

\appendix
\section{Supplementary data}
Supplementary material related to this article can be found online at \url{https://doi.org/10.1016/j.ascom.2022.100660}. Supplementary material contains: Configuration options set in the SLURM scripts to run the performance tests of the code.




\bibliography{bib_Gaia_AVU_GSR_Parallel}

\begin{thebibliography}{34}
\expandafter\ifx\csname natexlab\endcsname\relax\def\natexlab#1{#1}\fi
\providecommand{\url}[1]{\texttt{#1}}
\providecommand{\href}[2]{#2}
\providecommand{\path}[1]{#1}
\providecommand{\DOIprefix}{doi:}
\providecommand{\ArXivprefix}{arXiv:}
\providecommand{\URLprefix}{URL: }
\providecommand{\Pubmedprefix}{pmid:}
\providecommand{\doi}[1]{\href{http://dx.doi.org/#1}{\path{#1}}}
\providecommand{\Pubmed}[1]{\href{pmid:#1}{\path{#1}}}
\providecommand{\bibinfo}[2]{#2}
\ifx\xfnm\relax \def\xfnm[#1]{\unskip,\space#1}\fi
\bibitem[{{Aldinucci} et~al.(2021)}]{Aldinucci_2021}
\bibinfo{author}{{Aldinucci}, M.}, et~al., \bibinfo{year}{2021}.
\newblock \bibinfo{title}{Practical parallelization of scientific applications
  with openmp, openacc and mpi}.
\newblock \bibinfo{journal}{JPDC} \bibinfo{volume}{157},
  \bibinfo{pages}{13--29}.
\newblock \URLprefix
  \url{https://www.sciencedirect.com/science/article/pii/S0743731521001295},
  \DOIprefix\doi{https://doi.org/10.1016/j.jpdc.2021.05.017}.
\bibitem[{Amdahl(1967)}]{Amdahl_1967}
\bibinfo{author}{Amdahl, G.M.}, \bibinfo{year}{1967}.
\newblock \bibinfo{title}{Validity of the single processor approach to
  achieving large scale computing capabilities}, in:
  \bibinfo{booktitle}{Proceedings of the April 18-20, 1967, Spring Joint
  Computer Conference}, \bibinfo{publisher}{Association for Computing
  Machinery}, \bibinfo{address}{New York, NY, USA}. p.
  \bibinfo{pages}{483–485}.
\newblock \URLprefix \url{https://doi.org/10.1145/1465482.1465560},
  \DOIprefix\doi{10.1145/1465482.1465560}.
\bibitem[{Balay et~al.(2021a)Balay, Abhyankar, Adams, Brown, Brune, Buschelman,
  Dalcin, Dener, Eijkhout, Gropp, Kaushik, Knepley, May, McInnes, Mills,
  Munson, Rupp, Sanan, Smith, Zampini, Zhang and Zhang}]{petsc-user-ref}
\bibinfo{author}{Balay, S.}, \bibinfo{author}{Abhyankar, S.},
  \bibinfo{author}{Adams, M.F.}, \bibinfo{author}{Brown, J.},
  \bibinfo{author}{Brune, P.}, \bibinfo{author}{Buschelman, K.},
  \bibinfo{author}{Dalcin, L.}, \bibinfo{author}{Dener, A.},
  \bibinfo{author}{Eijkhout, V.}, \bibinfo{author}{Gropp, W.D.},
  \bibinfo{author}{Kaushik, D.}, \bibinfo{author}{Knepley, M.G.},
  \bibinfo{author}{May, D.A.}, \bibinfo{author}{McInnes, L.C.},
  \bibinfo{author}{Mills, R.T.}, \bibinfo{author}{Munson, T.},
  \bibinfo{author}{Rupp, K.}, \bibinfo{author}{Sanan, P.},
  \bibinfo{author}{Smith, B.F.}, \bibinfo{author}{Zampini, S.},
  \bibinfo{author}{Zhang, H.}, \bibinfo{author}{Zhang, H.},
  \bibinfo{year}{2021}a.
\newblock \bibinfo{title}{{PETS}c Users Manual}.
\newblock \bibinfo{type}{Technical Report} \bibinfo{number}{ANL-95/11 -
  Revision 3.15}. Argonne National Laboratory.
\bibitem[{Balay et~al.(2021b)Balay, Abhyankar, Adams, Brown, Brune, Buschelman,
  Dalcin, Dener, Eijkhout, Gropp, Kaushik, Knepley, May, McInnes, Mills,
  Munson, Rupp, Sanan, Smith, Zampini, Zhang and Zhang}]{petsc-web-page}
\bibinfo{author}{Balay, S.}, \bibinfo{author}{Abhyankar, S.},
  \bibinfo{author}{Adams, M.F.}, \bibinfo{author}{Brown, J.},
  \bibinfo{author}{Brune, P.}, \bibinfo{author}{Buschelman, K.},
  \bibinfo{author}{Dalcin, L.}, \bibinfo{author}{Dener, A.},
  \bibinfo{author}{Eijkhout, V.}, \bibinfo{author}{Gropp, W.D.},
  \bibinfo{author}{Kaushik, D.}, \bibinfo{author}{Knepley, M.G.},
  \bibinfo{author}{May, D.A.}, \bibinfo{author}{McInnes, L.C.},
  \bibinfo{author}{Mills, R.T.}, \bibinfo{author}{Munson, T.},
  \bibinfo{author}{Rupp, K.}, \bibinfo{author}{Sanan, P.},
  \bibinfo{author}{Smith, B.F.}, \bibinfo{author}{Zampini, S.},
  \bibinfo{author}{Zhang, H.}, \bibinfo{author}{Zhang, H.},
  \bibinfo{year}{2021}b.
\newblock \bibinfo{title}{{PETS}c {W}eb page}.
\newblock \bibinfo{note}{Https://petsc.org/}.
\bibitem[{Balay et~al.(1997)Balay, Gropp, McInnes and Smith}]{petsc-efficient}
\bibinfo{author}{Balay, S.}, \bibinfo{author}{Gropp, W.D.},
  \bibinfo{author}{McInnes, L.C.}, \bibinfo{author}{Smith, B.F.},
  \bibinfo{year}{1997}.
\newblock \bibinfo{title}{Efficient management of parallelism in object
  oriented numerical software libraries}, in: \bibinfo{editor}{Arge, E.},
  \bibinfo{editor}{Bruaset, A.M.}, \bibinfo{editor}{Langtangen, H.P.} (Eds.),
  \bibinfo{booktitle}{Modern Software Tools in Scientific Computing},
  \bibinfo{publisher}{Birkhauser Press}. pp. \bibinfo{pages}{163--202}.
\bibitem[{Baur and Austen(2005)}]{Baur_and_Austen_2005}
\bibinfo{author}{Baur, O.}, \bibinfo{author}{Austen, G.}, \bibinfo{year}{2005}.
\newblock \bibinfo{title}{A parallel iterative algorithm for large-scale
  problems of type potential field recovery from satellite data}, in:
  \bibinfo{booktitle}{Proceedings Joint CHAMP/GRACE Science Meeting,
  GeoForschungsZentrum Potsdam}.
\bibitem[{Becciani et~al.(2014)Becciani, Sciacca, Bandieramonte, Vecchiato,
  Bucciarelli and Lattanzi}]{Becciani_2014}
\bibinfo{author}{Becciani, U.}, \bibinfo{author}{Sciacca, E.},
  \bibinfo{author}{Bandieramonte, M.}, \bibinfo{author}{Vecchiato, A.},
  \bibinfo{author}{Bucciarelli, B.}, \bibinfo{author}{Lattanzi, M.G.},
  \bibinfo{year}{2014}.
\newblock \bibinfo{title}{Solving a very large-scale sparse linear system with
  a parallel algorithm in the gaia mission}, in: \bibinfo{booktitle}{2014
  International Conference on High Performance Computing Simulation (HPCS)},
  pp. \bibinfo{pages}{104--111}.
\newblock \DOIprefix\doi{10.1109/HPCSim.2014.6903675}.
\bibitem[{{Bertone} et~al.(2017){Bertone}, {Vecchiato}, {Bucciarelli},
  {Crosta}, {Lattanzi}, {Bianchi}, {Angonin} and {Le
  Poncin-Lafitte}}]{Bertone_2017}
\bibinfo{author}{{Bertone}, S.}, \bibinfo{author}{{Vecchiato}, A.},
  \bibinfo{author}{{Bucciarelli}, B.}, \bibinfo{author}{{Crosta}, M.},
  \bibinfo{author}{{Lattanzi}, M.G.}, \bibinfo{author}{{Bianchi}, L.},
  \bibinfo{author}{{Angonin}, M.C.}, \bibinfo{author}{{Le Poncin-Lafitte}, C.},
  \bibinfo{year}{2017}.
\newblock \bibinfo{title}{{Application of time transfer functions to Gaia's
  global astrometry. Validation on DPAC simulated Gaia-like observations}}.
\newblock \bibinfo{journal}{A\&A} \bibinfo{volume}{608}, \bibinfo{pages}{A83}.
\newblock \DOIprefix\doi{10.1051/0004-6361/201731654},
  \href{http://arxiv.org/abs/1708.00541}{\tt arXiv:1708.00541}.
\bibitem[{Bin et~al.(2020)Bin, Wu, Shao, Zhou and Bin}]{Bin_2020}
\bibinfo{author}{Bin, G.}, \bibinfo{author}{Wu, S.}, \bibinfo{author}{Shao,
  M.}, \bibinfo{author}{Zhou, Z.}, \bibinfo{author}{Bin, G.},
  \bibinfo{year}{2020}.
\newblock \bibinfo{title}{Irn-mlsqr: An improved iterative reweight norm
  approach to the inverse problem of electrocardiography incorporating
  factorization-free preconditioned lsqr}.
\newblock \bibinfo{journal}{J. Electrocardiol.} \bibinfo{volume}{62},
  \bibinfo{pages}{190--199}.
\newblock \URLprefix
  \url{https://www.sciencedirect.com/science/article/pii/S0022073620305379},
  \DOIprefix\doi{https://doi.org/10.1016/j.jelectrocard.2020.08.017}.
\bibitem[{Borriello et~al.(1986)Borriello, Dalessandro, Murgolo and
  Prezioso}]{Borriello_1986}
\bibinfo{author}{Borriello, L.}, \bibinfo{author}{Dalessandro, F.},
  \bibinfo{author}{Murgolo, F.}, \bibinfo{author}{Prezioso, G.},
  \bibinfo{year}{1986}.
\newblock \bibinfo{title}{Hipparcos-the reduction chain of observations and
  double star recognition using an image processing approach}.
\newblock \bibinfo{journal}{Mem. Soc. Astron. Ital.} \bibinfo{volume}{57},
  \bibinfo{pages}{267--289}.
\bibitem[{Brown et~al.(2021)Brown, Vallenari, Prusti, de~Bruijne, Babusiaux,
  Biermann, Creevey, Evans, Eyer and et~al.}]{Gaia_EDR3_2021}
\bibinfo{author}{Brown, A.G.A.}, \bibinfo{author}{Vallenari, A.},
  \bibinfo{author}{Prusti, T.}, \bibinfo{author}{de~Bruijne, J.H.J.},
  \bibinfo{author}{Babusiaux, C.}, \bibinfo{author}{Biermann, M.},
  \bibinfo{author}{Creevey, O.L.}, \bibinfo{author}{Evans, D.W.},
  \bibinfo{author}{Eyer, L.}, \bibinfo{author}{et~al.}, \bibinfo{year}{2021}.
\newblock \bibinfo{title}{Gaia early data release 3}.
\newblock \bibinfo{journal}{A\&A} \bibinfo{volume}{650}, \bibinfo{pages}{C3}.
\newblock \URLprefix \url{http://dx.doi.org/10.1051/0004-6361/202039657e},
  \DOIprefix\doi{10.1051/0004-6361/202039657e}.
\bibitem[{{Crosta} et~al.(2017){Crosta}, {Geralico}, {Lattanzi} and
  {Vecchiato}}]{Crosta_2017}
\bibinfo{author}{{Crosta}, M.}, \bibinfo{author}{{Geralico}, A.},
  \bibinfo{author}{{Lattanzi}, M.G.}, \bibinfo{author}{{Vecchiato}, A.},
  \bibinfo{year}{2017}.
\newblock \bibinfo{title}{{General relativistic observable for gravitational
  astrometry in the context of the Gaia mission and beyond}}.
\newblock \bibinfo{journal}{\prd} \bibinfo{volume}{96},
  \bibinfo{pages}{104030}.
\newblock \DOIprefix\doi{10.1103/PhysRevD.96.104030}.
\bibitem[{Flores et~al.(2016)Flores, Vidal and Verd{\'u}}]{Flores_2016}
\bibinfo{author}{Flores, L.}, \bibinfo{author}{Vidal, V.},
  \bibinfo{author}{Verd{\'u}, G.}, \bibinfo{year}{2016}.
\newblock \bibinfo{title}{Gpu based algorithms in ct imaging}.
\newblock \bibinfo{journal}{AMGP} \bibinfo{volume}{3}, \bibinfo{pages}{25--31}.
\bibitem[{Guo et~al.(2021)Guo, Zhao, Yu, He, He and Song}]{Guo_2021}
\bibinfo{author}{Guo, H.}, \bibinfo{author}{Zhao, H.}, \bibinfo{author}{Yu,
  J.}, \bibinfo{author}{He, X.}, \bibinfo{author}{He, X.},
  \bibinfo{author}{Song, X.}, \bibinfo{year}{2021}.
\newblock \bibinfo{title}{X-ray luminescence computed tomography using a hybrid
  proton propagation model and lasso-lsqr algorithm}.
\newblock \bibinfo{journal}{J. Biophotonics} , \bibinfo{pages}{e202100089}.
\bibitem[{Gustafson(1988)}]{Gustafson_1988}
\bibinfo{author}{Gustafson, J.L.}, \bibinfo{year}{1988}.
\newblock \bibinfo{title}{Reevaluating amdahl's law}.
\newblock \bibinfo{journal}{Commun. ACM} \bibinfo{volume}{31},
  \bibinfo{pages}{532–533}.
\newblock \URLprefix \url{https://doi.org/10.1145/42411.42415},
  \DOIprefix\doi{10.1145/42411.42415}.
\bibitem[{{Hees} et~al.(2018){Hees}, {Le Poncin-Lafitte}, {Hestroffer} and
  {David}}]{Hees_2018}
\bibinfo{author}{{Hees}, A.}, \bibinfo{author}{{Le Poncin-Lafitte}, C.},
  \bibinfo{author}{{Hestroffer}, D.}, \bibinfo{author}{{David}, P.},
  \bibinfo{year}{2018}.
\newblock \bibinfo{title}{{Local tests of gravitation with Gaia observations of
  Solar System Objects}}, in: \bibinfo{editor}{{Recio-Blanco}, A.},
  \bibinfo{editor}{{de Laverny}, P.}, \bibinfo{editor}{{Brown}, A.G.A.},
  \bibinfo{editor}{{Prusti}, T.} (Eds.), \bibinfo{booktitle}{Astrometry and
  Astrophysics in the Gaia Sky}, pp. \bibinfo{pages}{63--66}.
\newblock \DOIprefix\doi{10.1017/S1743921317005907},
  \href{http://arxiv.org/abs/1709.05329}{\tt arXiv:1709.05329}.
\bibitem[{Huang et~al.(2013)Huang, Dennis, Wang and Chen}]{Huang_2013}
\bibinfo{author}{Huang, H.}, \bibinfo{author}{Dennis, J.M.},
  \bibinfo{author}{Wang, L.}, \bibinfo{author}{Chen, P.}, \bibinfo{year}{2013}.
\newblock \bibinfo{title}{A scalable parallel lsqr algorithm for solving
  large-scale linear system for tomographic problems: a case study in seismic
  tomography}.
\newblock \bibinfo{journal}{Procedia Comput. Sci.} \bibinfo{volume}{18},
  \bibinfo{pages}{581--590}.
\bibitem[{Huang et~al.(2012)Huang, Wang, Lee and Chen}]{Huang_2012}
\bibinfo{author}{Huang, H.}, \bibinfo{author}{Wang, L.}, \bibinfo{author}{Lee,
  E.J.}, \bibinfo{author}{Chen, P.}, \bibinfo{year}{2012}.
\newblock \bibinfo{title}{An mpi-cuda implementation and optimization for
  parallel sparse equations and least squares (lsqr)}.
\newblock \bibinfo{journal}{Procedia Comput. Sci.} \bibinfo{volume}{9},
  \bibinfo{pages}{76--85}.
\bibitem[{Jaffri et~al.(2020)Jaffri, Shi, Abrar, Ahmad and Yang}]{Jaffri_2020}
\bibinfo{author}{Jaffri, N.R.}, \bibinfo{author}{Shi, L.},
  \bibinfo{author}{Abrar, U.}, \bibinfo{author}{Ahmad, A.},
  \bibinfo{author}{Yang, J.}, \bibinfo{year}{2020}.
\newblock \bibinfo{title}{Electrical resistance tomographic image enhancement
  using mrnsd and lsqr}, in: \bibinfo{booktitle}{Proceedings of the 2020 5th
  International Conference on Multimedia Systems and Signal Processing}, pp.
  \bibinfo{pages}{16--20}.
\bibitem[{Joulidehsar et~al.(2018)Joulidehsar, Moradzadeh and
  Ardejani}]{Joulidehsar_2018}
\bibinfo{author}{Joulidehsar, F.}, \bibinfo{author}{Moradzadeh, A.},
  \bibinfo{author}{Ardejani, F.D.}, \bibinfo{year}{2018}.
\newblock \bibinfo{title}{An improved 3d joint inversion method of potential
  field data using cross-gradient constraint and lsqr method}.
\newblock \bibinfo{journal}{Pure Appl. Geophys.} \bibinfo{volume}{175},
  \bibinfo{pages}{4389--4409}.
\bibitem[{{Krolikowski} et~al.(2021){Krolikowski}, {Kraus} and
  {Rizzuto}}]{Krolikowski_2021}
\bibinfo{author}{{Krolikowski}, D.M.}, \bibinfo{author}{{Kraus}, A.L.},
  \bibinfo{author}{{Rizzuto}, A.C.}, \bibinfo{year}{2021}.
\newblock \bibinfo{title}{{Gaia EDR3 Reveals the Substructure and Complicated
  Star Formation History of the Greater Taurus-Auriga Star-forming Complex}}.
\newblock \bibinfo{journal}{\aj} \bibinfo{volume}{162}, \bibinfo{pages}{110}.
\newblock \DOIprefix\doi{10.3847/1538-3881/ac0632},
  \href{http://arxiv.org/abs/2105.13370}{\tt arXiv:2105.13370}.
\bibitem[{Liang et~al.(2019a)Liang, Jiao, Fan and Yang}]{LIANG_2019}
\bibinfo{author}{Liang, S.X.}, \bibinfo{author}{Jiao, Y.J.},
  \bibinfo{author}{Fan, W.X.}, \bibinfo{author}{Yang, B.Z.},
  \bibinfo{year}{2019}a.
\newblock \bibinfo{title}{3d inversion of magnetic data based on lsqr method
  and correlation coefficient self constrained}.
\newblock \bibinfo{journal}{Progress in Geophysics} \bibinfo{volume}{34},
  \bibinfo{pages}{1475--1480}.
\newblock \DOIprefix\doi{https://doi.org/10.6038/pg2019CC0275}.
\bibitem[{Liang et~al.(2019b)Liang, Wang, Jiao, Liao and
  Jing}]{LSQR_geology_2019}
\bibinfo{author}{Liang, S.X.}, \bibinfo{author}{Wang, Q.},
  \bibinfo{author}{Jiao, Y.J.}, \bibinfo{author}{Liao, G.Z.},
  \bibinfo{author}{Jing, G.}, \bibinfo{year}{2019}b.
\newblock \bibinfo{title}{Lsqr - analysis and evaluation of the potential field
  inversion using lsqr method}.
\newblock \bibinfo{journal}{Geophysical and Geochemical Exploration}
  \bibinfo{volume}{43}, \bibinfo{pages}{359--366}.
\newblock \DOIprefix\doi{https://doi.org/10.11720/wtyht.2019.1261}.
\bibitem[{Ling et~al.(2019)Ling, Jia, Lu and Yang}]{Ling_2019}
\bibinfo{author}{Ling, S.T.}, \bibinfo{author}{Jia, Z.G.}, \bibinfo{author}{Lu,
  X.}, \bibinfo{author}{Yang, B.}, \bibinfo{year}{2019}.
\newblock \bibinfo{title}{Matrix lsqr algorithm for structured solutions to
  quaternionic least squares problem}.
\newblock \bibinfo{journal}{Comput. Math. Appl.} \bibinfo{volume}{77},
  \bibinfo{pages}{830--845}.
\newblock \URLprefix
  \url{https://www.sciencedirect.com/science/article/pii/S0898122118306205},
  \DOIprefix\doi{https://doi.org/10.1016/j.camwa.2018.10.023}.
\bibitem[{Liu et~al.(2006)Liu, Liu, Liu and Hao}]{Liu_2006}
\bibinfo{author}{Liu, J.S.}, \bibinfo{author}{Liu, F.T.}, \bibinfo{author}{Liu,
  J.}, \bibinfo{author}{Hao, T.Y.}, \bibinfo{year}{2006}.
\newblock \bibinfo{title}{Parallel lsqr algorithms used in seismic tomography}.
\newblock \bibinfo{journal}{Chin. J. Geophys.} \bibinfo{volume}{49},
  \bibinfo{pages}{483--488}.
\bibitem[{Naghibzadeh and van~der Veen(2017)}]{Naghibzadeh_and_vanderVeen_2017}
\bibinfo{author}{Naghibzadeh, S.}, \bibinfo{author}{van~der Veen, A.J.},
  \bibinfo{year}{2017}.
\newblock \bibinfo{title}{Radioastronomical least squares image reconstruction
  with iteration regularized krylov subspaces and beamforming-based prior
  conditioning}, in: \bibinfo{booktitle}{2017 IEEE International Conference on
  Acoustics, Speech and Signal Processing (ICASSP)},
  \bibinfo{organization}{IEEE}. pp. \bibinfo{pages}{3385--3389}.
\bibitem[{Paige and Saunders(1982a)}]{Paige_and_Saunders_1982a}
\bibinfo{author}{Paige, C.C.}, \bibinfo{author}{Saunders, M.A.},
  \bibinfo{year}{1982a}.
\newblock \bibinfo{title}{Lsqr: An algorithm for sparse linear equations and
  sparse least squares}.
\newblock \bibinfo{journal}{ACM Trans. Math. Softw. (TOMS)}
  \bibinfo{volume}{8}, \bibinfo{pages}{43--71}.
\bibitem[{Paige and Saunders(1982b)}]{Paige_and_Saunders_1982b}
\bibinfo{author}{Paige, C.C.}, \bibinfo{author}{Saunders, M.A.},
  \bibinfo{year}{1982b}.
\newblock \bibinfo{title}{Algorithm 583: Lsqr: Sparse linear equations and
  least squares problems}.
\newblock \bibinfo{journal}{ACM Trans. Math. Softw. (TOMS)}
  \bibinfo{volume}{8}, \bibinfo{pages}{195--209}.
\bibitem[{Penghui and Houbiao(2020)}]{Penghui_2020}
\bibinfo{author}{Penghui, H.}, \bibinfo{author}{Houbiao, L.},
  \bibinfo{year}{2020}.
\newblock \bibinfo{title}{A note on the least squares qr (lsqr) algorithm}.
\newblock \bibinfo{journal}{Math. Numer. Sin.} \bibinfo{volume}{42},
  \bibinfo{pages}{487--496}.
\bibitem[{Reichel and Ye(2008)}]{Reichel_and_Ye_2008}
\bibinfo{author}{Reichel, L.}, \bibinfo{author}{Ye, Q.}, \bibinfo{year}{2008}.
\newblock \bibinfo{title}{A generalized lsqr algorithm}.
\newblock \bibinfo{journal}{Numer. Linear Algebra Appl.} \bibinfo{volume}{15},
  \bibinfo{pages}{643--660}.
\bibitem[{{Van der Marel}(1988)}]{VanderMarel_1988}
\bibinfo{author}{{Van der Marel}, H.}, \bibinfo{year}{1988}.
\newblock \bibinfo{title}{{On the ``great circle reduction'' in the data
  analysis for the astrometric satellite HIPPARCOS''}}.
\newblock Ph.D. thesis. Delft University of Technology, Netherlands.
\bibitem[{{Vecchiato} et~al.(2018){Vecchiato}, {Bucciarelli}, {Lattanzi},
  {Becciani}, {Bianchi}, {Abbas}, {Sciacca}, {Messineo} and {De
  March}}]{Vecchiato_2018}
\bibinfo{author}{{Vecchiato}, A.}, \bibinfo{author}{{Bucciarelli}, B.},
  \bibinfo{author}{{Lattanzi}, M.G.}, \bibinfo{author}{{Becciani}, U.},
  \bibinfo{author}{{Bianchi}, L.}, \bibinfo{author}{{Abbas}, U.},
  \bibinfo{author}{{Sciacca}, E.}, \bibinfo{author}{{Messineo}, R.},
  \bibinfo{author}{{De March}, R.}, \bibinfo{year}{2018}.
\newblock \bibinfo{title}{{The global sphere reconstruction (GSR).
  Demonstrating an independent implementation of the astrometric core solution
  for Gaia}}.
\newblock \bibinfo{journal}{A\&A} \bibinfo{volume}{620}, \bibinfo{pages}{A40}.
\newblock \DOIprefix\doi{10.1051/0004-6361/201833254},
  \href{http://arxiv.org/abs/1809.05145}{\tt arXiv:1809.05145}.
\bibitem[{{Vecchiato} et~al.(2003){Vecchiato}, {Lattanzi}, {Bucciarelli},
  {Crosta}, {de Felice} and {Gai}}]{Vecchiato_2003}
\bibinfo{author}{{Vecchiato}, A.}, \bibinfo{author}{{Lattanzi}, M.G.},
  \bibinfo{author}{{Bucciarelli}, B.}, \bibinfo{author}{{Crosta}, M.},
  \bibinfo{author}{{de Felice}, F.}, \bibinfo{author}{{Gai}, M.},
  \bibinfo{year}{2003}.
\newblock \bibinfo{title}{{Testing general relativity by micro-arcsecond global
  astrometry}}.
\newblock \bibinfo{journal}{A\&A} \bibinfo{volume}{399},
  \bibinfo{pages}{337--342}.
\newblock \DOIprefix\doi{10.1051/0004-6361:20021785},
  \href{http://arxiv.org/abs/astro-ph/0301323}{\tt arXiv:astro-ph/0301323}.
\bibitem[{Yoo et~al.(2011)Yoo, Baker, Pearce and Henson}]{Yoo_2011}
\bibinfo{author}{Yoo, A.}, \bibinfo{author}{Baker, A.H.},
  \bibinfo{author}{Pearce, R.}, \bibinfo{author}{Henson, V.E.},
  \bibinfo{year}{2011}.
\newblock \bibinfo{title}{A scalable eigensolver for large scale-free graphs
  using 2d graph partitioning}, in: \bibinfo{booktitle}{Proceedings of 2011
  International Conference for High Performance Computing, Networking, Storage
  and Analysis}, \bibinfo{publisher}{Association for Computing Machinery},
  \bibinfo{address}{New York, NY, USA}.
\newblock \URLprefix \url{https://doi.org/10.1145/2063384.2063469},
  \DOIprefix\doi{10.1145/2063384.2063469}.

\end{thebibliography}

\end{document}